\documentclass[11pt,a4paper]{article}
\usepackage[latin1]{inputenc} 
\usepackage[british]{babel}
\usepackage[fleqn]{amsmath}
\usepackage{latexsym}
\usepackage[dvips]{graphics,graphicx,psfrag}
\usepackage{setspace}
\usepackage[compat2]{geometry}
\usepackage{multirow}
\usepackage{times}
\usepackage{hyperref}
\usepackage[round,comma,authoryear,sectionbib]{natbib}

\usepackage{placeins}
\graphicspath{{./}}

\usepackage{articlestyle}
\long\def\symbolfootnote[#1]#2{\begingroup%
\def\thefootnote{\fnsymbol{footnote}}\footnote[#1]{#2}\endgroup}

\begin{document}

\section*{\textsf{\LARGE
How do Markov approximations compare with other \\ methods for large
spatial data sets?
}}

{\Large David Bolin$^{a,}$\symbolfootnote[1]{Corresponding author. Tel.: +46 46 2227974; fax: +46 46 2224623; \textit{Email address:} \textsf{bolin@maths.lth.se} (David Bolin)}\, Finn Lindgren$^{b}$}\\

\noindent
\textit{$^{a}$Mathematical Statistics,
Centre for Mathematical Sciences,
Lund University, Sweden}\\
\noindent
\textit{$^{b}$Department of Mathematical Sciences,
Norwegian University of Science and Technology,
Trondheim, Norway}\\

\begin{center}
\line(1,0){475}
\end{center}
\textbf{Abstract}\\
The Mat\'{e}rn covariance function is a popular choice for modeling
dependence in spatial environmental data. Standard Mat\'{e}rn
covariance models are, however, often computationally infeasible for
large data sets. In this work, recent results for Markov
approximations of Gaussian Mat\'{e}rn fields based on Hilbert space
approximations are extended using wavelet basis functions.  These
Markov approximations are compared with two of the most popular
methods for efficient covariance approximations; covariance tapering
and the process convolution method.  The results show that, for a given
computational cost, the Markov methods have a substantial gain in
accuracy compared with the other methods.\\

\noindent
\textit{Key words:} Mat\'{e}rn covariances, Kriging, Wavelets, Markov random
fields, Covariance tapering, process convolutions, Computational efficiency

\begin{center}
\line(1,0){475}
\end{center}

\section{Introduction}\label{paperB:sec:intro}
The traditional methods in spatial statistics were typically developed
without any considerations of computational efficiency. In many of the
classical applications of spatial statistics in environmental
sciences, the cost for obtaining measurements limited the size of the
data sets to ranges where computational cost was not an issue. Today,
however, with the increasing use of remote sensing satellites,
producing many large climate data sets, computational efficiency is
often a crucial property.

In recent decades, several techniques for building computationally
efficient models have been suggested. In many of these techniques, the
main assumption is that a latent, zero mean Gaussian process
$X(\mv{s})$ can be expressed, or at least approximated, through some
finite basis expansion
\begin{equation}\label{paperB:lowrank_proc}
X(\mv{s}) = \sum_{j=1}^n w_j\basis_j(\mv{s}),
\end{equation}
where $w_j$ are Gaussian random variables, and $\{\basis_j\}_{j=1}^{n}$
are pre-defined basis functions. The justification for using
these basis expansions is usually that they converge to the true spatial
model as $n$ tends to infinity. However, for a finite $n$,
the choice of the weights and basis functions will greatly affect the
approximation error and the computational efficiency of the
model. Hence, if one wants an accurate model for a given computational
cost, asymptotic arguments are insufficient.\par

If the process $X(\mv{s})$ has a discrete spectral density, one can
obtain an approximation on the form \eqref{paperB:lowrank_proc} by
truncating the spectral expansion of the process. Another way to
obtain an, in some sense optimal, expansion on the form
\eqref{paperB:lowrank_proc} is to use the the eigenfunctions of the
covariance function for the latent field $X(\mv{s})$ as a basis, which
is usually called the Karhunen-Lo\`{e}ve (KL) transform. The problem
with the KL transform is that analytic expressions for the
eigenfunctions are only known in a few simple cases, which are often
insufficient to represent the covariance structure in real data
sets. Numerical approximations of the eigenfunctions can be obtained
for a given covariance function; however, the covariance function is
in most cases not known, but has to be estimated from data. In these
cases, it is infeasible to use the KL expansion in the parameter
estimation, which is often the most computationally demanding part of
the analysis. The spectral representation has a similar problem since
the computationally efficient methods are usually restricted to
stationary models with gridded data, and are not applicable in more
general situations.  Thus, to be useful for a broad range of practical
applications, the methods should be applicable to a wide family of
stationary covariance functions, and be extendable to nonstationary
covariance structures.\par

One method that fulfills these requirements is the
process convolution approach \citep{barry96,higdon01,cressie02,rodriges10}. In
this method, the stochastic field, $X(\mv{s})$, is defined as the
convolution of a Gaussian white noise process with some convolution
kernel $k(\mv{s})$. This convolution is then approximated with a sum on the
form \eqref{paperB:lowrank_proc} to get a discrete model representation.
Process convolution approximations are computationally efficient if a small number of basis functions can be used, but in practice, this will often give a
poor approximation of the continuous convolution model.

A popular method for creating computationally efficient approximations is
covariance tapering \citep{furrer06}. This method can not be written as an approximation on the form \eqref{paperB:lowrank_proc}, but the idea is
instead to taper the true covariance to zero beyond a certain range by
multiplying the covariance function with some compactly supported
taper function \citep{gneiting02}. This facilitates the use of sparse matrix techniques that increases the computational efficiency, at the cost of replacing
the original model with a different model, which can lead to problems
depending on the spatial structure of the data locations. However,
the method is applicable to both stationary and nonstationary
covariance models, and instead of choosing the set of basis functions in
\eqref{paperB:lowrank_proc}, the taper range and the taper function
has to be chosen.

\cite{nychka02} used a wavelet basis in the expansion
\eqref{paperB:lowrank_proc}, and showed that by allowing for some
correlation among the random variables $w_j$, one gets a flexible
model that can be used for estimating nonstationary covariance
structures. As a motivating example, they showed that using a wavelet
basis, computationally efficient approximations to the popular
Mat\'{e}rn covariance functions can be obtained using only a few
nonzero correlations for the weights $w_j$. The approximations were,
however, obtained numerically, and no explicit representations were
derived.

\citet{Rue02} showed that general stationary covariance models can be
closely approximated by Markov random fields, by numerically
minimizing the error in the resulting covariances.  \cite{song08}
extended the method by applying different loss criteria, such as
minimizing the spectral error or the Kullback-Leibler divergence.  A
drawback of the methods is that, just as for the KL and wavelet
approaches, the numerical optimisation must in general be performed
for each distinct parameter configuration.

Recently, \citet{lindgren07} derived an explicit method for producing
computationally efficient approximations to the Mat\'{e}rn covariance
family. The method uses the fact that a random process on $\R^d$ with
a Mat\'{e}rn covariance function is a solution to a certain stochastic
partial differential equation (SPDE). By considering weak solutions to
this SPDE with respect to some set of local basis functions
$\{\basis_j\}_{j=1}^n$, an approximation on the form
\eqref{paperB:lowrank_proc} is obtained, where the stochastic weights
have a sparse precision matrix (inverse covariance matrix), that can
be written directly as a function of the parameters, without any need
for costly numerical calculations. The method is also extendable to more general stationary and nonstationary models by extending the generating SPDE \citep{lindgren10, bolin09b}.

In this paper, we use methods from \cite{lindgren07} and
\cite{lindgren10} to algebraically compute the weights $w_j$ for
wavelet based approximations to Gaussian Mat\'{e}rn fields
(Section~\ref{paperB:sec:SPDE}). For certain wavelet bases, the
weights form a Gaussian Markov Random Field (GMRF), which greatly
increases the computational efficiency of the approximation. For other
wavelet bases, such as the one used in \cite{nychka02}, the weights
can be well approximated with a GMRF. 

In order to evaluate the practical usefulness of the different
approaches, a detailed analysis of the computational aspects of the
spatial prediction problem is performed
(Section~\ref{paperB:sec:kriging} and
Section~\ref{paperB:sec:comparison}).
The results show that the GMRF methods are more efficient and accurate
than both the process convolution approach and the covariance tapering
method.

\section{Spatial prediction and computational cost}\label{paperB:sec:kriging}
As a motivating example for why computational efficiency is important, 
 consider spatial prediction. The most widely used method for
spatial prediction is commonly known as linear kriging in
geostatistics.
Let $Y(\mv{s})$ be an observation of a latent Gaussian field, $X(\mv{s})$,
under mean zero Gaussian measurement noise, $\mathcal{E}(\mv{s})$, uncorrelated with $X$ and with some covariance function $r_{\mathcal{E}}(\mv{s},\mv{t})$,
\begin{equation}
  \label{paperB:eq:measurement}
  Y(\mv{s}) = X(\mv{s}) + \mathcal{E}(\mv{s}),
\end{equation}
and let $\mu(\mv{s})$ and $r(\mv{s},\mv{t})$ be the mean value
function and covariance function for $X(\mv{s})$
respectively. Depending on the assumptions on $\mu(\mv{s})$, linear
kriging is usually divided into simple kriging (if $\mu$ is known),
ordinary kriging (if $\mu$ is unknown but independent of $\mv{s}$),
and universal kriging (if $\mu$ is unknown and can be expressed as a
linear combination of some deterministic basis functions). To limit
the scope of this article, parameter estimation will not be
considered, and to simplify the notations, we let $\mu(\mv{s})\equiv
0$. It should, however, be noted that all results in later sections
regarding computational efficiency also hold in the cases of ordinary
kriging and universal kriging. For more details on kriging, see e.g.\cite{stein99} or \cite{schabenberger05}.

Let $r(\mv{s},\mv{t})$ have some parametric structure, and let the vector
$\mv{\gamma}$ contain all covariance parameters. Let $\mv{Y}$ be a
vector containing the observations, $\mv{X}_1$ be a vector containing
$X(\mv{s})$ evaluated at the measurement locations, $\mv{s}_1, \ldots,
\mv{s}_m$, and let $\mv{X}_2$ be a vector containing $X(\mv{s})$ at
the locations, $\hat{\mv{s}}_1, \ldots, \hat{\mv{s}}_{\hat{m}}$, for
which the kriging predictor should be calculated.
With $\mv{X} = (\mv{X}_1^{\trsp},\mv{X}_2^{\trsp})^{\trsp}$, one has
$\mv{X}_1 = \mv{A}_1\mv{X}$, and $\mv{X}_2 = \mv{A}_2\mv{X}$ for two
diagonal matrices $\mv{A}_1$ and $\mv{A}_2$, and the model can now be written
as
\begin{align*}
  \mv{X}|\mv{\gamma} &\sim \pN(\mv{0}, \mv{\Sigma}_{X}),\\
  \mv{Y}|\mv{X} &\sim \pN(\mv{A}_1\mv{X},\mv{\Sigma}_{\mathcal{E}}),
\end{align*}
where $\mv{\Sigma}_{X}$ is the covariance matrix for $\mv{X}$ and $\mv{\Sigma}_{\mathcal{E}}$ contains the covariances $r_{\mathcal{E}}(\mv{s}_i,\mv{s}_j)$
It is straightforward to show that $\mv{X}|\mv{Y},\mv{\gamma} \sim \pN(\hat{\mv{\Sigma}}\mv{A}_1\mv{\Sigma}_{\mathcal{E}}^{-1}\mv{Y},\hat{\mv{\Sigma}})$, where $\hat{\mv{\Sigma}} =
(\mv{\Sigma}_X^{-1}+\mv{A}_1^{\trsp}\mv{\Sigma}_{\mathcal{E}}^{-1}\mv{A}_1)^{-1}$,
and the well known expression for the kriging predictor is now given by
the conditional mean
\begin{align}
 \pE(\mv{X}_2|\mv{Y},\mv{\gamma}) & =  \mv{A}_2\hat{\mv{\Sigma}}\mv{A}_1\mv{\Sigma}_{\mathcal{E}}^{-1}\mv{Y}\notag
  =
 \mv{A}_2\mv{\Sigma}_{X}\mv{A}_1^{\trsp}(\mv{A}_1\mv{\Sigma}_X\mv{A}_1^{\trsp}+\mv{\Sigma}_{\mathcal{E}})^{-1}\mv{Y}\notag\\
 & =  \mv{\Sigma}_{X_2
   X_1}(\mv{\Sigma}_{X_1}+\mv{\Sigma}_{\mathcal{E}})^{-1}\mv{Y}
  =  \mv{\Sigma}_{X_2 X_1}\mv{\Sigma}_{Y}^{-1}\mv{Y},\label{paperB:eq:kriging}
\end{align}
where the elements on row $i$ and column $j$ in $\mv{\Sigma}_{X_2 X_1}$
and $\mv{\Sigma}_{Y}$ are given by the covariances
$r(\hat{\mv{s}}_i,\mv{s}_j)$ and
$r(\mv{s}_i,\mv{s}_j) + r_{\mathcal{E}}(\mv{s}_i,\mv{s}_j)$ respectively. To get the standard
expression for the variance of the kriging predictor, the Woodbury
identity is used on $\hat{\mv{\Sigma}}$:
\begin{align*}
 \pV(\mv{X}_2|\mv{Y},\mv{\gamma}) & = \mv{A}_2(\mv{\Sigma}_X^{-1}+\mv{A}_1^{\trsp}\mv{\Sigma}_{\mathcal{E}}^{-1}\mv{A}_1)^{-1}\mv{A}_2^{\trsp} \\
 &=  \mv{A}_2\mv{\Sigma}_{X}\mv{A}_2 -
 \mv{A}_2\mv{\Sigma}_X\mv{A}_1^{\trsp}(\mv{A}_1\mv{\Sigma}_X\mv{A}_1^{\trsp}+\mv{\Sigma}_{\mathcal{E}})\mv{A}_1\mv{\Sigma}_X\mv{A}_2^{\trsp} \\
 &= \mv{\Sigma}_{X_2} - \mv{\Sigma}_{X_2
   X_1}\mv{\Sigma}_Y^{-1}\mv{\Sigma}_{X_2 X_1}^{\trsp}.
\end{align*}
If there are no simplifying assumptions on $\mv{\Sigma}_{X}$, the
computational cost for calculating the kriging predictor is
$\Ordo(\hat{m}m+m^3)$, and the cost for calculating the variance is
even higher. This means that with $1000$ measurements, the number of
operations needed for the kriging prediction for a single location is
on the order of $10^9$. These computations are thus not feasible for a
large data set where one might have more than $10^6$
measurements. 

The methods described in Section \ref{paperB:sec:intro} all make different
approximations in order to reduce the computational cost for
calculating the kriging predictor and its variance. These different
approximations, and their impact on the computational cost, are
described in more detail in Section \ref{paperB:sec:comparison};
however, to get a general idea of how the computational efficiency can
be increased, consider the kriging predictor for a model on the form
\eqref{paperB:lowrank_proc}. The field $\mv{X}$ can then be written as
$\mv{X} = \Basis\mv{w} \sim \pN(\mv{0}, \Basis\mv{\Sigma}_w\Basis^{\trsp})$,
where column $i$ in the matrix $\Basis$ contains the basis function
$\basis_i(\mv{s})$ evaluated at all measurement locations and all locations where the kriging prediction is to be calculated. Let $\Basis_1 = \mv{A}_1\Basis$ and $\Basis_2 = \mv{A}_2\Basis$ be the matrices containing the basis
functions evaluated at the measurement locations and the kriging
locations respectively. The kriging predictor is then
\begin{equation}\label{paperB:lowrank_krig}
 \pE(\mv{X}_2|\mv{Y},\mv{\gamma}) = \Basis_2 (\mv{\Sigma}_w^{-1} + \Basis_1^{\trsp}\mv{\Sigma}_{\mathcal{E}}^{-1}\Basis_1)^{-1}\Basis_1\mv{\Sigma}_{\mathcal{E}}^{-1}\mv{Y}.
\end{equation}
If the measurement noise is Gaussian white noise,
$\mv{\Sigma}_{\mathcal{E}}$ is diagonal and easy to invert. If
$\mv{\Sigma}_w^{-1}$ is either known, or easy to calculate, the most expensive
calculation in \eqref{paperB:lowrank_krig} is to solve $\mv{u} =
(\mv{\Sigma}_w^{-1} + \Basis_1^{\trsp}\mv{\Sigma}_{\mathcal{E}}^{-1}
\Basis_1)^{-1}\Basis_1\mv{\Sigma}_{\mathcal{E}}^{-1}\mv{Y}$.
This is a linear system of $n$ equations, where $n$ is the number of
basis functions used in the approximation. Thus, the easiest way of
reducing the computational cost is to choose $n \ll m$, which is what
is done in the convolution approach. Another approach
is to ensure that $ (\mv{\Sigma}_w^{-1} + \Basis_1^{\trsp}
\mv{\Sigma}_{\mathcal{E}}^{-1} \Basis_1)$ is a sparse
matrix. Sparse matrix techniques can then be used to calculate the kriging
predictor, and the computational cost can be reduced
without reducing the number of basis functions in the approximation.
If a wavelet basis is used, $ \Basis_1^{\trsp}
\mv{\Sigma}_{\mathcal{E}}^{-1} \Basis_1$ will be sparse, and in
Section~\ref{paperB:sec:SPDE}, it is shown that the precision matrix
$\mv{Q}_w = \mv{\Sigma}_w^{-1}$ can also be chosen as a sparse matrix
by using the Hilbert space approximation technique by
\cite{lindgren10}.

\section{Wavelet approximations}\label{paperB:sec:SPDE}
In the remainder of this paper, the focus is on the family of Mat\'{e}rn
covariance functions \citep{matern60} and the computational efficiency
of some different techniques for approximating Gaussian Mat\'{e}rn
fields. This section shows how wavelet bases can be used in
the Hilbert space approximation technique by \cite{lindgren10} to
obtain computationally efficient Mat\'{e}rn approximations.

\subsection{The Mat\'{e}rn covariance family}
Because of its versatility, the Mat\'{e}rn covariance family is the
most popular choice for modeling spatial data~\citep{stein99}. There are a few different parameterizations of the Mat\'{e}rn covariance function in the literature, and the one most suitable in our context is
\begin{equation}\label{paperB:eq:matern}
r(\mv{h}) = \frac{2^{1-\nu}\phi^2}{(4\pi)^{\frac{d}{2}}\Gamma(\nu + \frac{d}{2})\kappa^{2\nu}}(\kappa\|\mv{h}\|)^{\nu}K_{\nu}(\kappa\|\mv{h}\|),
\end{equation}
where $\nu$ is a shape parameter, $\kappa^2$ a scale parameter, $\phi^2$ a variance parameter, and $K_{\nu}$ is a modified Bessel function of the second kind of order $\nu>0$. With this parametrization, the variance of a field
with this covariance is $r(\mv{0}) = \phi^2\Gamma(\nu)(4\pi)^{-\frac{d}{2}}\Gamma(\nu+\frac{d}{2})^{-1}\kappa^{-2\nu}$, and the associated spectral density is
\begin{equation}\label{paperB:matern_spec}
S(\mv{\omega}) = \frac{\phi^2}{(2\pi)^d}\frac{1}{(\kappa^2 + \|\mv{\omega}\|^2)^{\nu + \frac{d}{2}}}.
\end{equation}
For the special case $\nu=0.5$, the Mat\'{e}rn covariance function is
the exponential covariance function. The smoothness of the field
increases with $\nu$, and in the limit as $\nu
\rightarrow \infty$, the covariance function is a Gaussian
covariance function if $\kappa$ is also scaled accordingly, which gives an infinitely differentiable field.

\subsection{Hilbert space approximations}\label{paperB:sec:galerkin}
As noted by \cite{whittle63}, a random process with the
covariance \eqref{paperB:eq:matern} is a solution to the SPDE
\begin{equation}\label{paperB:sde}
(\kappa^2-\Delta)^{\frac{\alpha}{2}}X(\mv{s}) = \phi \noise(\mv{s}),
\end{equation}
where $\noise(\mv{s})$ is Gaussian white noise, $\Delta$ is the Laplacian,
and $\alpha = \nu + d/2$.  The key idea in \citet{lindgren10}
is to approximate the solution to the SPDE using a basis expansion on the form \eqref{paperB:lowrank_proc}. The starting point of the approximation is to consider the stochastic weak formulation of the SPDE
\begin{equation}\label{paperB:eq:weak}
\left\{\scal{b_i}{(\kappa^2-\Delta)^{\frac{\alpha}{2}}X},i=1,\ldots,n_b\right\} 
\overset{d}{=} 
\left\{\scal{b_i}{\phi\noise},i=1,\ldots,n_b\right\}.
\end{equation}
Here $\overset{d}{=}$ denotes equality in distribution, $\scal{f}{g} =
\int f(\mv{s})g(\mv{s})\md \mv{s}$, and equality should hold for every finite set of test functions $\{b_i,i=1,\ldots,n_b\}$ from some appropriate space. A finite element approximation of the solution $X$ is then obtained by representing it as a finite basis expansion on the form \eqref{paperB:lowrank_proc}, where the stochastic weights are calculated by requiring \eqref{paperB:eq:weak} to hold for only a specific set of test functions $\{b_i,i=1,\ldots,n\}$ and $\{\basis_i\}$ is a set of predetermined basis functions. We illustrate the more general results from \citet{lindgren10} with
the special case $\alpha=2$, where one uses $b_i=\basis_i$ and one then has
\begin{equation}
   \scal{\basis_i}{(\kappa^2-\Delta)X} = \sum_{j=1}^n w_j\scal{\basis_i}{(\kappa^2-\Delta)\basis_j}.
\end{equation} 
By introducing the matrix $\mv{K}$ with elements $\mv{K}_{i,j} =
\scal{\basis_i}{(\kappa^2-\Delta)\basis_j}$ and the vector
$\mv{w} = (w_1, \ldots, w_n)^{\trsp}$, the left hand side of
\eqref{paperB:eq:weak} can be written as $\mv{K}\mv{w}$.
Since, by Lemma 1 in \cite{lindgren10}
\begin{equation*}
  \scal{\basis_i}{(\kappa^2-\Delta)\basis_j} = \kappa^2\scal{\basis_i}{\basis_j}-\scal{\basis_i}{\Delta\basis_j} =\kappa^2\scal{\basis_i}{\basis_j}+\scal{\nabla\basis_i}{\nabla\basis_j},
\end{equation*}
the matrix $\mv{K}$ can be written as the sum $\mv{K} = \kappa^2\mv{C} +
\mv{G}$ where $\mv{C}_{i,j} = \scal{\basis_i}{\basis_j}$
and $\mv{G}_{i,j} = \scal{\nabla\basis_i}{\nabla\basis_j}$. The right hand side of \eqref{paperB:eq:weak} can be shown to be Gaussian
with mean zero and covariance $\phi^2\mv{C}$ and thus get that $\mv{w} \sim \pN(0,\phi^2\mv{K}^{-1}\mv{C}\mv{K}^{-1})$.

For the second fundamental case, $\alpha=1$, \cite{lindgren10} show that $\mv{w} \sim \pN(\mv{0},\phi^2\mv{K}^{-1})$ and for higher order $\alpha \in \N$, the weak solution is obtained recursively using these two fundamental cases.
For example, if $\alpha = 4$ the solution to
$(\kappa^2 - \Delta)^2X_0(\mv{s}) = \phi \noise(\mv{s})$ is obtained by solving
$(\kappa^2 -\Delta)X_0(\mv{s}) = \tilde{X}(\mv{s})$, where $\tilde{X}$ is the solution for the case $\alpha = 2$. This results in a precision matrix for the weights $\mv{Q}_{\alpha}$ defined recursively as 
\begin{equation}\label{eq:Qa}
\mv{Q}_{\alpha} =\mv{K}\mv{C}^{-1}\mv{Q}_{\alpha-2}\mv{C}^{-1}\mv{K}, \quad \alpha=3,4,\ldots
\end{equation}
where $\mv{Q}_1 = \phi^{-2}\mv{K}$ and $\mv{Q}_2 = \phi^{-2}\mv{K}^{\trsp}\mv{C}^{-1}\mv{K}$.
Thus, all Mat\'{e}rn fields with $\nu+d/2 \in \N$ can be approximated
through this procedure. For more details, see \citet{lindgren07} and
\citet{lindgren10}.  The results from~\citet{Rue02} show that accurate
Markov approximations exist also for other $\nu$-values, and one approximate approach to finding explicit expressions for such models was given in the authors' response in \citet{lindgren10}.
However, in many practical applications $\nu$ cannot be estimated
reliably~\citep{zhang04}, and using only a discrete set of $\nu$-values is
not necessarily a significant restriction.

\subsection{Wavelet basis functions}
In the previous section, nothing was said about how the the basis functions
$\{\basis_i\}$ should be chosen. The following sections, however,
shows that wavelet bases have many desirable properties which makes
them suitable to use in the Hilbert space approximations on $\R^d$. In
this section, a brief introduction to multiresolution analysis and
wavelets is given.\par

A multiresolution analysis on $\R$ is a sequence of closed
approximation subspaces $\{V_j\}_{j\in \Z}$ of functions in
$L^{2}(\R)$ such that $V_j \subset V_{j+1}$, $\closure{\bigcup_{j\in \Z}V_j} = L^{2}(\R)$, and $\bigcap_{j\in \Z}V_j = \{0\}$, where $\closure$ is the closure, and $f(s)\in V_j$ if and only if
$f(2^{-j}s) \in V_0$. This last requirement is the multiresolution
requirement because this implies that all the approximation spaces
$V_j$ are scaled versions of the space $V_0$. A multiresolution
analysis is generated starting with a function usually called a father
function or a scaling function. The function $\varphi\in L^2(\R)$ is
called a scaling function for $\{V_j\}_{j\in \Z}$ if it satisfies the two-scale relation
\begin{equation}\label{paperB:scaling}
  \varphi(s) = \sum_{k\in \Z}p_k\varphi(2s-k),
\end{equation}
for some square-summable sequence $\{p_k\}_{k\in\Z}$ and the translates $\{\varphi(s-k)\}_{k\in\Z}$ form an orthonormal basis for $V_0$. Given the
multiresolution analysis $\{V_j\}_{j\in \Z}$, the wavelet spaces
$\{W_j\}_{j\in\Z}$ are then defined as the orthogonal complements of
$V_j$ in $V_{j+1}$ for each $j$, and one can show that $W_j$ is the
span of $\{\psi(2^j s - k)\}_{k\in\Z}$, where the wavelet $\psi$ is
defined as $\psi(s) = \sum_{k\in\Z}(-1)^k\overline{p_{1-k}}\varphi(2s-k)$.

Given the spaces $W_j$, $V_j$ can be decomposed as the direct sum
\begin{equation}\label{paperB:eq:multires}
  V_j = V_0 \oplus W_0  \oplus W_1  \oplus \ldots  \oplus W_{j-1}.
\end{equation}
Several choices of scaling functions have been presented in the
literature. Among the most widely used constructions are the B-spline
wavelets \citep{Chui92} and the Daubechies wavelets \citep{daub92} that both have several desirable properties for our purposes.

The scaling function of B-spline wavelets are $m$:th order B-splines
with knots at the integers. Because of this, there exists closed form expressions for the corresponding wavelets, and the wavelets have compact support since the $m$:th order scaling function has support on $(0,m+1)$.
The wavelets are orthogonal at different scales, but translates at the
same scale are not orthogonal. This property is usually referred to as
semi-orthogonality.

\begin{figure}[t]
\begin{center}
%
%
\begin{psfrags}%
\psfragscanon%
%
\psfrag{s04}[b][b]{\setlength{\tabcolsep}{0pt}\begin{tabular}{c}Scaling function\end{tabular}}%
%
\psfrag{x01}[t][t]{0}%
\psfrag{x02}[t][t]{1}%
\psfrag{x03}[t][t]{2}%
\psfrag{x04}[t][t]{3}%
\psfrag{x05}[t][t]{4}%
\psfrag{x06}[t][t]{5}%
%
\psfrag{v01}[r][r]{-0.4}%
\psfrag{v02}[r][r]{0}%
\psfrag{v03}[r][r]{0.4}%
\psfrag{v04}[r][r]{0.8}%
\psfrag{v05}[r][r]{1.2}%
%
\resizebox{6cm}{!}{\includegraphics{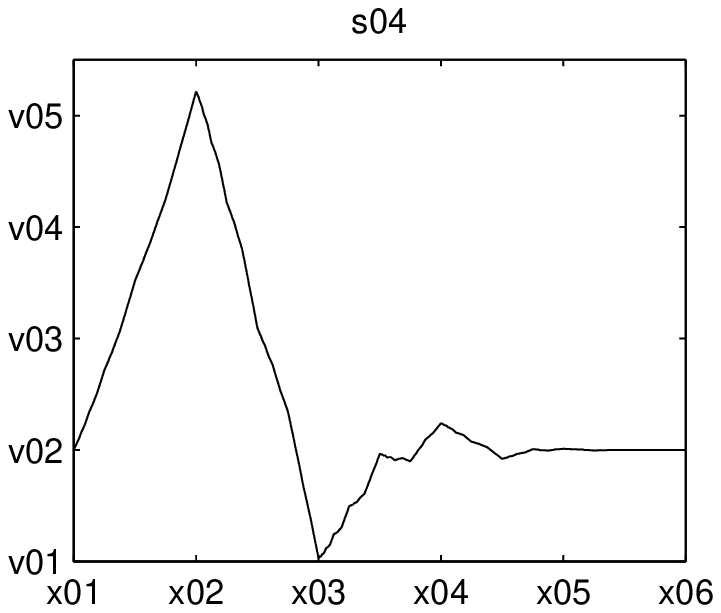}}%
\end{psfrags}%
%
%
%
%
\begin{psfrags}%
\psfragscanon%
%
\psfrag{s04}[b][b]{\setlength{\tabcolsep}{0pt}\begin{tabular}{c}Wavelet\end{tabular}}%
%
\psfrag{x01}[t][t]{-2}%
\psfrag{x02}[t][t]{-1}%
\psfrag{x03}[t][t]{0}%
\psfrag{x04}[t][t]{1}%
\psfrag{x05}[t][t]{2}%
\psfrag{x06}[t][t]{3}%
%
\psfrag{v01}[r][r]{-1.2}%
\psfrag{v02}[r][r]{-0.6}%
\psfrag{v03}[r][r]{0}%
\psfrag{v04}[r][r]{0.6}%
\psfrag{v05}[r][r]{1.2}%
\psfrag{v06}[r][r]{1.8}%
%
\resizebox{6cm}{!}{\includegraphics{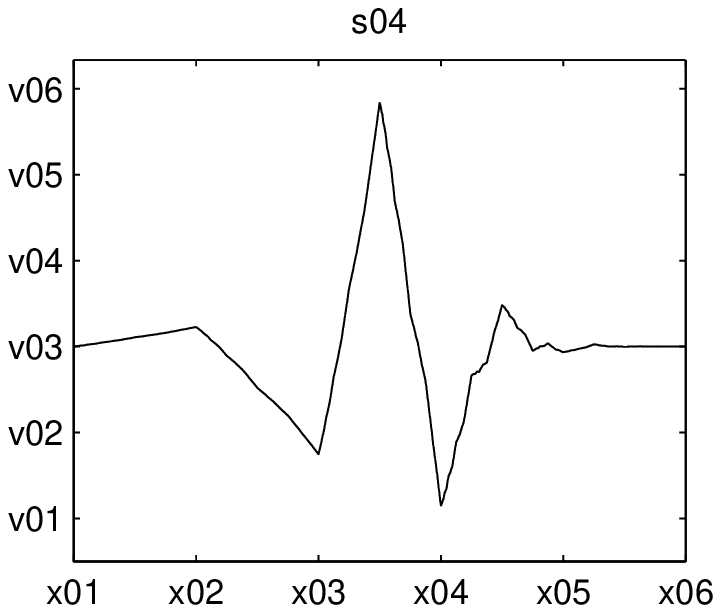}}%
\end{psfrags}%
%
%
\end{center}
\caption{The DB3 scaling function and wavelet.} 
\label{paperB:fig:DB3}
\end{figure}

The Daubechies wavelets form a hierarchy of compactly supported
orthogonal wavelets that are constructed to have the highest number of
vanishing moments for a given support width. This generates a
family of wavelets with an increasing degree of smoothness. Except for the first Daubechies wavelet, there are no closed form expressions for these wavelets; however, for practical purposes, this is not a problem because the exact values for the wavelets at dyadic points can be obtained very fast using the Cascade algorithm \citep{burrus88}. In this work, the DB3 wavelet is used because it is the first wavelet in the family that has one continuous derivative. The DB3
wavelet and its scaling function are shown in Figure \ref{paperB:fig:DB3}.

\subsection{Explicit wavelet Hilbert space approximations}
To use the Hilbert space approximation for a given basis, the precision matrix for the weights $\mv{Q}_{\alpha}$ has to be calculated. By \eqref{eq:Qa}, we only have to be able to calculate the matrices $\mv{C}$ and $\mv{G}$ to built the precision matrix for any $\alpha\in\N$. The elements in
these matrices are inner products between the basis functions:
\begin{align}
\label{paperB:eq:CandGcalculation}
\mv{C}_{i,j} &= \int\basis_i(\mv{s})\basis_j(\mv{s})\md \mv{s}, & 
  \mv{G}_{i,j} &= \int(\nabla\basis_i(\mv{s}))^{\trsp}\nabla\basis_j(\mv{s})\md \mv{s}.
\end{align}
This section shows how these elements can be calculated for the DB3
wavelets and the B-spline wavelets. When using a wavelet basis in
practice, one always have to choose a finest scale, $J$, to work with.
Given that the subspace $V_J$ is used as an approximation of
$L^2(\R)$, one can use two different bases. Either one works with the
direct basis for $V_J$, that consists of scaled and translated versions of the
father function $\varphi(s)$, or one can use the multiresolution
decomposition \eqref{paperB:eq:multires}. In what follows, both these
cases are considered.

\subsubsection{Daubechies wavelets on $\R$}
For the Daubechies wavelets, the matrix $\mv{C}$ is the identity
matrix since these wavelets form an orthonormal basis for
$L^2(\R)$. Thus, only the matrix $\mv{G}$ has to be calculated. If the
direct basis for $V_J$ is used, $\mv{G}$ contains inner products on
the form
\begin{equation}\label{paperB:db_gip}
  \scal{\nabla \varphi(2^J s-k)}{\nabla
    \varphi(2^J s-l)} =  2^{J}\scal{\nabla \varphi(s)}{\nabla
    \varphi(s-l+k)} \equiv 2^{J}\Lambda(k-l).
\end{equation}
Because the scaling function has compact support on $[0,2N-1]$, these inner
products are only non-zero if $k-l\in [-(2N-2),2N-2]$. Thus, the
matrix $\mv{G}$ is sparse, which implies that the weights
$\mv{w}$ in \eqref{paperB:lowrank_proc} form a GMRF. Since there are
no closed form expressions for the Daubechies wavelets, there is no
hope in finding a closed form expression for the non-zero inner
products \eqref{paperB:db_gip}. Furthermore, standard numerical
quadrature for calculating the inner products is too inaccurate due to
the highly oscillating nature of the gradients. However, utilizing
properties of the wavelets, one can calculate an approximation of the
inner product of arbitrary precision by solving a system of linear
equations. It is outside the scope of this paper to present the full
method, but the basic principle is to construct a system of linear
equations by using the scaling- and moment equations for the
wavelets. This system is then solved using, for example, LU
factorization. For details, see \cite{latto91}.

Using this technique for the DB3 wavelets, the following nonzero values for
$\Lambda(\eta)$ are obtained
\begin{align*}
  \Lambda(0) &= 5.267, & \Lambda(\pm1) &= -3.390, & \Lambda(\pm2) &=
  0.876, \\
  \Lambda(\pm3) &= -0.114, & \Lambda(\pm4) &= -0.00535.
\end{align*}
These values are calculated once and tabulated for constructing
the $\mv{G}$ matrix, which is a band matrix with $2^{J}\Lambda(0)$ on the main diagonal, $2^{J}\Lambda(1)$ on the first off diagonals, et cetera.

\begin{figure}[t]
\begin{center}
%
%
\begin{psfrags}%
\psfragscanon%
%
\psfrag{s03}[t][t]{\setlength{\tabcolsep}{0pt}\begin{tabular}{c}nz = 55798\end{tabular}}%
\psfrag{s04}[b][b]{\setlength{\tabcolsep}{0pt}\begin{tabular}{c}Multiresolution basis\end{tabular}}%
%
\psfrag{x01}[t][t]{0}%
\psfrag{x02}[t][t]{200}%
\psfrag{x03}[t][t]{400}%
\psfrag{x04}[t][t]{600}%
\psfrag{x05}[t][t]{800}%
%
\psfrag{v01}[r][r]{0}%
\psfrag{v02}[r][r]{200}%
\psfrag{v03}[r][r]{400}%
\psfrag{v04}[r][r]{600}%
\psfrag{v05}[r][r]{800}%
%
\resizebox{6cm}{!}{\includegraphics{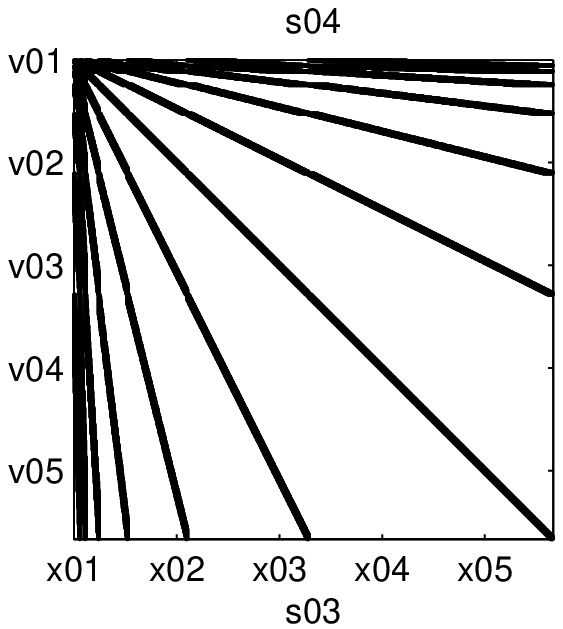}}%
\end{psfrags}%
%
%
%
%
\begin{psfrags}%
\psfragscanon%
%
\psfrag{s03}[t][t]{\setlength{\tabcolsep}{0pt}\begin{tabular}{c}nz = 8368\end{tabular}}%
\psfrag{s04}[b][b]{\setlength{\tabcolsep}{0pt}\begin{tabular}{c}Direct basis\end{tabular}}%
%
\psfrag{x01}[t][t]{0}%
\psfrag{x02}[t][t]{200}%
\psfrag{x03}[t][t]{400}%
\psfrag{x04}[t][t]{600}%
\psfrag{x05}[t][t]{800}%
%
\psfrag{v01}[r][r]{0}%
\psfrag{v02}[r][r]{200}%
\psfrag{v03}[r][r]{400}%
\psfrag{v04}[r][r]{600}%
\psfrag{v05}[r][r]{800}%
%
\resizebox{6cm}{!}{\includegraphics{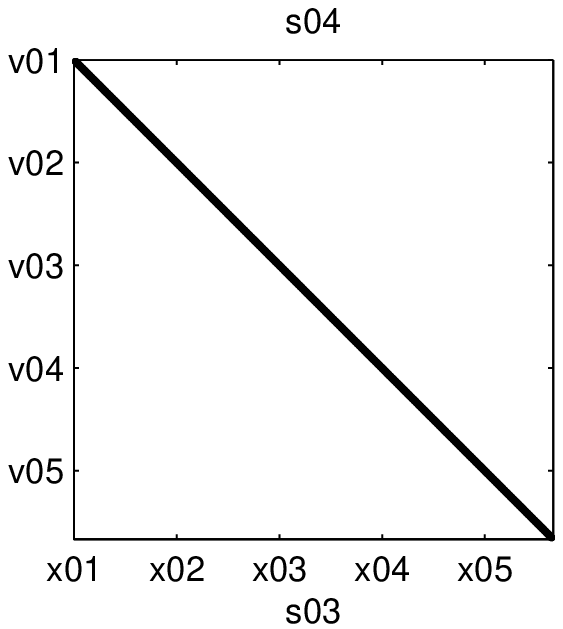}}%
\end{psfrags}%
%
%
\end{center}
\caption{The non-zero elements in the $\mv{G}$ matrices for a multiresolution
  DB3 basis with five layer of wavelets and the corresponding direct
  basis. $6.4\%$ of the elements are non-zero for the multiresolution basis whereas only $0.96\%$ of the elements are non-zero for the direct basis.}
\label{paperB:fig:spyG}
\end{figure}

If the multiresolution decomposition \eqref{paperB:eq:multires} is
used as a basis for $V_J$, one also needs the inner products
\begin{equation*}
\scal{\nabla \psi(2^j s-k)}{\nabla   \psi(2^i s-l)}, \quad i,j\in\Z.  
\end{equation*}
Because of the two-scale relation \eqref{paperB:scaling}, every wavelet $\psi(2^j s-k)$ can be
written as a finite sum of the scaling function at scale $J$. Using
this property, the $\mv{G}$ matrix can be constructed efficiently using only the already computed $\Lambda$ values. Figure~\ref{paperB:fig:spyG} shows the structure of the $\mv{G}$ matrices for a multiresolution DB3 basis with five layers of wavelets and the corresponding direct basis. Note that there are fewer
non-zero elements in the precision matrix for the direct basis. Hence,
it is more computationally efficient to use the direct basis instead
of the multiresolution basis.

\subsubsection{B-spline wavelets on $\R$}
For the B-spline wavelets, the matrices $\mv{C}$ and $\mv{G}$ can be
calculated directly from the closed form expressions for the basis
functions and their derivatives.
When a direct basis is used on $\R$, $\mv{C}$ is a band matrix with
bandwidth $m+1$, if the $m$:th order spline wavelet is used.
For example, for $m=1$, calculating \eqref{paperB:eq:CandGcalculation}
gives
\begin{align*}
\mv{C}_{i,j} &= 2^{-J} \cdot
\begin{cases}
2/3, & i=j, \\
1/6, & |i-j|=1, \\
0& \text{otherwise,}
\end{cases}
&
\mv{G}_{i,j} &= 2^{J} \cdot
\begin{cases}
2, & i=j, \\
-1, & |i-j|=1, \\
0& \text{otherwise.}
\end{cases}
\end{align*}
Since the expression for the precision matrix for the weights $\mv{w}$
contains the inverse of $\mv{C}$, it is a dense matrix. Hence,
$\mv{C}^{-1}$ has to be approximated with a sparse matrix if $\mv{Q}$
should be sparse.  This issue is addressed in \cite{lindgren10} by
lowering the integration order of $\scal{\basis_i}{\basis_j}$, which
results in an approximate, diagonal $\mv{C}$ matrix, $\tilde{\mv{C}}$,
with diagonal elements $\tilde{\mv{C}}_{ii} =
\sum_{k=1}^n\mv{C}_{ik}$. In Section \ref{paperB:sec:comparison}, the
effect of this approximation on the covariance approximation for the
basis expansion is studied in some detail.
For the multiresolution basis, the matrices are block diagonal, and
this approximation is not applicable.

\subsubsection{Wavelets on $\R^d$}
The easiest way of constructing a wavelet basis for $L^2(\R^d)$ is to
use the tensor product functions generated by $d$ one-dimensional
wavelet bases. Let $\varphi$ be the scaling function
for a multiresolution on $\R$, the father function can be
written as $\bar{\varphi}(x_1, \ldots, x_d) =
\prod_{i=1}^d\varphi(x_i)$. The scalar product
$\scal{\nabla\bar{\varphi}(\mv{x})}{
  \nabla\bar{\varphi}(\mv{x}+\mv{\eta})}$, where $\mv{\eta}$
now is a multi-integer shift in $d$ dimensions, can then be written as
\begin{align*}
 \hspace{-0.5cm} \scal{\nabla\bar{\varphi}(\mv{x})}{
    \nabla\bar{\varphi}(\mv{x}+\mv{\eta})} &=
  \scal{\nabla \prod_{i=1}^d \varphi(x)}{
    \nabla  \prod_{i=1}^d \varphi(x+\eta_i)} \\
  &= \sum_{i=1}^d\int_{\R^d} \frac{\pd \varphi(x_i)}{\pd x_i}
  \frac{\pd \varphi(x_i+\eta_i)}{\pd x_i}  \prod_{j\neq i} \varphi(x_j)
     \varphi(x_j+\eta_j) \md\mv{x} \\
     &=  \sum_{i=1}^d\Lambda(\eta_i)  \prod_{j\neq i}\int_{\R} \varphi(x_j)
     \varphi(x_j+\eta_j) \md x_j .
\end{align*}
This expression looks rather complicated, but it implies a very
simple Kronecker structure for $\mv{G}_d$, the $\mv{G}$ matrix in $\R^d$. For
example, in $\R^2$ and $\R^3$,
\begin{align*}
  \mv{G}_2 &= \mv{G}_1\otimes\mv{C}_1 + \mv{C}_1\otimes\mv{G}_1\\
  \mv{G}_3 &= \mv{G}_1\otimes\mv{C}_1\otimes\mv{C}_1 +
  \mv{C}_1\otimes\mv{G}_1\otimes\mv{C}_1 + \mv{C}_1\otimes\mv{C}_1\otimes\mv{G}_1,
\end{align*}
where $\mv{G}_1$ and $\mv{C}_1$ are the $\mv{G}$ and $\mv{C}$ matrices
for the corresponding one-dimensional basis and $\otimes$ denotes the
Kronecker product. Similarly, $\mv{C}_2 = \mv{C}_1\otimes\mv{C}_1$, and $\mv{C}_3 = \mv{C}_1\otimes\mv{C}_1\otimes\mv{C}_1$. These expressions hold both if the direct basis for $V_J$ if used or if the multiresolution construction
\eqref{paperB:eq:multires} is used for the one-dimensional spaces.
For Daubechies wavelets, the $\mv{C}$ matrix is the identity matrix
for all $d\geq 1$.  This also holds for the direct B-spline basis if
the diagonal approximation is used for $\mv{C}_1$.

\section{Comparison}\label{paperB:sec:comparison}

As discussed in Section \ref{paperB:sec:kriging} is computational efficiency often an important aspect in practical applications. However, the computation time for obtaining for example an approximate kriging prediction is in itself not that interesting unless one also knows how accurate it is. We will therefore in this section compare the wavelet Markov approximations with two other popular methods, covariance tapering and process convolutions, with respect to their accuracy and computationally efficiency when used for kriging.  

Before the comparison, we give a brief introduction to the process convolution method and the covariance tapering method and discuss the methods' computational properties. As mentioned in Section \ref{paperB:sec:kriging}, the computational
cost for the kriging prediction for a single location based on $m$ observations is $\Ordo(m^3)$. In what follows, the corresponding computational costs for the three different approximation methods are presented. We start with the wavelet Markov approximations and then look at the process convolutions and the covariance tapering method. After this, an initial comparison of the different wavelet approximations is performed in Section \ref{sec:covcomp} and then the full kriging comparison is presented in Section \ref{sec:krigingcomp}-\ref{sec:tapercomp}.

\subsection{Wavelet approximations}
When using a wavelet basis, one can either work with the direct basis for the approximation space $V_J$ or do the wavelet decomposition into the direct sum of
$J-1$ wavelet spaces and $V_0$. If one only is interested in the
approximation error, the decomposition into wavelet spaces is not
necessary and it is more efficient to work in the direct basis for $V_J$ since this will result in a precision matrix with fewer nonzero elements. Therefore we only use the direct bases for $V_J$ in the comparisons in this section. 

The wavelet approximations are on the form \eqref{paperB:lowrank_proc}, so Equation \eqref{paperB:lowrank_krig} is used to calculate the kriging predictor. However, since an explicit expression for the precision matrix for the weights $\mv{w}$ exists for this method, we rewrite the equation as
\begin{equation*}
 \pE(\mv{X}_2|\mv{Y},\mv{\gamma}) = \Basis_2 (\mv{Q}_w + \Basis_1^{\trsp}\mv{Q}_{\mathcal{E}}\Basis_1)^{-1}\Basis_1\mv{Q}_{\mathcal{E}}\mv{Y},
\end{equation*}
where $\mv{Q}_{\mathcal{E}} = \mv{\Sigma}_{\mathcal{E}}^{-1}$ is diagonal if $\mathcal{E}$ is Gaussian white noise.
If the number of kriging locations is small, the computationally
demanding step is again to solve a system on the form
\begin{equation*}
 \mv{u} = (\mv{Q}_w + \Basis_1^{\trsp}\mv{Q}_{\mathcal{E}}\Basis_1)^{-1}\mv{v}.
\end{equation*}
Now, if the Daubechies wavelets or the Markov approximated spline
wavelets are used, both $\mv{Q}_w$ and
$\Basis_1^{\trsp}\mv{Q}_{\mathcal{E}}\Basis_1$ are sparse and
positive definite matrices. The system is therefore most efficiently
solved using Cholesky factorization, forward substitution, and back
substitution. The forward substitution and back substitution are much faster than calculating the Cholesky triangle $\mv{L}$, so the computational
complexity for the kriging predictor is determined by the calculation
of $\mv{L}$.  Because of the sparsity structure, this computational
cost is in general $\Ordo(n)$, $\Ordo(n^{3/2})$, and $\Ordo(n^2)$ for
problems in one, two, and three dimensions respectively
\cite[see][]{rue1}. If the spline bases are used without the markov approximation, the computational cost instead is $O(n^3)$ since $\mv{Q}_w$ then is dense. It should be noted here that any basis could be used in the SPDE approximation, but in order to get good computational properties we need both $\mv{Q}_w$ and $\Basis_1^{\trsp}\mv{Q}_{\mathcal{E}}\Basis_1$ to be sparse. This is the reason for why for example Fourier bases are not appropriate to use in the SPDE formulation since $B_1$ in this case always is a dense matrix.  

\subsection{Process convolutions}\label{conv_fields_intro}
In the process convolution method, the Gaussian random field $X(\mv{s})$ on
$\R^d$ is specified as a process convolution
\begin{equation}\label{paperB:convdef}
X(\mv{s}) = \int k(\mv{s},\mv{u}) \mathcal{B}(\md\mv{u}),
\end{equation}
where $k$ is some deterministic kernel function and $\mathcal{B}$ is a Brownian sheet. One of the advantages with this construction is that nonstationary
fields easily are constructed by allowing the convolution kernel to be
dependent on location. If, however, the process is stationary we have $k(\mv{s},\mv{u}) = k(\mv{s}-\mv{u})$ and the covariance function for $X$ is
$r(\mv{\tau}) = \int k(\mv{u}-\mv{\tau})k(\mv{u})\md\mv{u}$.
Thus, the covariance function and the kernel $k$ are related through
\begin{equation*}
k = \mathcal{F}^{-1}\left(\frac{1}{(2\pi)^{\frac{d}{2}}}\sqrt{\mathcal{F}(r)}\right) = \mathcal{F}^{-1}\left(\frac{1}{(2\pi)^{\frac{d}{2}}} \sqrt{S}\right),
\end{equation*}
where $S$ is the spectral density for $X(\mv{s})$ and $\mathcal{F}$ denotes the Fourier transform \citep{higdon01}. Since the spectral density for a Mat\'{e}rn covariance function in dimension $d$ with parameters $\nu$, $\phi^2$, and $\kappa$ is given by \eqref{paperB:matern_spec}, one finds that the corresponding kernel is a Mat\'{e}rn covariance function with parameters $\nu_k = \frac{\nu}{2}-\frac{d}{4}$, $\phi_k^2
= \phi$, and $\kappa_k = \kappa$.\par
An approximation of \eqref{paperB:convdef} which is commonly used in convolution based modeling is
\begin{equation*}
  X(\mv{s}) \approx \sum_{j=1}^{n}k(\mv{s}-\mv{u}_j)w_j,
\end{equation*}
where $\mv{u}_1,\ldots, \mv{u}_n$ are some fixed locations in the
domain, and $w_j$ are independent zero mean Gaussian variables with
variances equal to the area associated with each $\mv{u}_j$. Thus,
this approximation is on the form \eqref{paperB:lowrank_proc}, with
basis functions $\xi_j(\mv{s})=k(\mv{s}-\mv{u}_j)$. When this approximation is used, Equation \eqref{paperB:lowrank_krig} is
used to calculate the kriging predictor. Because the basis functions
in the expansion are Mat\'{e}rn covariance functions, the matrices
$\Basis_1$ and $\Basis_2$ are dense. Thus, even though
both $\mv{\Sigma}_{\mathcal{E}}$ and $\mv{\Sigma}_{w}^{-1}$ are
diagonal matrices, one still has to solve a system on the form
\begin{equation*}
 \mv{u} = (\mv{\Sigma}_w^{-1} + \Basis_1^{\trsp}\mv{\Sigma}_{\mathcal{E}}^{-1}\Basis_1)^{-1}\mv{v}
\end{equation*}
where $(\mv{\Sigma}_w^{-1} +
\Basis_1^{\trsp}\mv{\Sigma}_{\mathcal{E}}^{-1}\Basis_1)$ is a dense $n$ by $n$ matrix. The computational cost for both constructing and inverting the matrix is
$\Ordo(mn^2+n^3)$, where $n$ is the number of basis functions used in the
basis expansion. For kriging prediction of $\hat{m}$ locations, the total
computational complexity is $\Ordo(\hat{m}n+mn^2+n^3)$.

\subsection{Covariance tapering}\label{taper_intro}
Covariance tapering is not a method for constructing covariance
models, but a method for approximating a given covariance model to
increase the computational efficiency. The idea is simply
to to taper the true covariance, $r(\mv{\tau})$, to zero beyond a certain
range, $\theta$, by multiplying the covariance function with some
compactly supported positive definite taper function $r_{\theta}(\mv{\tau})$. Using the tapered covariance,
\begin{equation*}
  r_{tap}(\mv{\tau}) = r_{\theta}(\mv{\tau})r(\mv{\tau}),
\end{equation*}
the matrix $\mv{\Sigma}_Y$ in the expression for the kriging predictor
\eqref{paperB:eq:kriging} is sparse, which facilitates the use of
sparse matrix techniques that increases the computational
efficiency. The taper function should, of course, also be chosen such that the basic shape of the true covariance function is preserved, and of especial importance for asymptotic considerations is that the smoothness at the origin is preserved.

\cite{furrer06} studied the accuracy and numerical efficiency of tapered
Mat\'{e}rn covariance functions, and to be able to compare their results to Mat\'{e}rn approximations obtained by the wavelet Hilbert space approximations and the process convolution method, we use their choice of taper functions:
\begin{align*}
&\hspace{-0.3cm}\mbox{Wendland$_{1}$:} & &\hspace{-0.3cm}r_{\theta}(\mv{\tau}) =
  \left(\max\left[1-\frac{\|\mv{\tau}\|}{\theta},0\right]\right)^4\left(1+4\frac{\|\mv{\tau}\|}{\theta}\right),\\
      &\hspace{-0.3cm}\mbox{Wendland$_{2}$:} & &\hspace{-0.3cm}r_{\theta}(\mv{\tau}) =
  \left(\max\left[1-\frac{\|\mv{\tau}\|}{\theta},0\right]\right)^6\left(1+6\frac{\|\mv{\tau}\|}{\theta}+\frac{35\|\mv{\tau}\|^2}{2\theta^2}\right).
\end{align*}
These taper functions were first introduced by \citet{wendland95}.
For dimension $d\leq3$, the Wendland$_{1}$ function is a valid taper function for the Mat\'{e}rn covariance function if $\nu\leq 1.5$, and the Wendland$_{2}$ functions is a valid taper function if $\nu\leq 2.5$. \cite{furrer06} found that
Wendland$_{1}$ was slightly better than Wendland$_{2}$ for a given
$\nu$, so we use Wendland$_{1}$ for all cases when $\nu\leq 1.5$
and Wendland$_{2}$ if $1.5<\nu \leq 2.5$.

If a tapered Mat\'{e}rn covariance is used, the kriging predictor can
be written as
\begin{align*}
 \pE(\mv{X}_2|\mv{Y},\mv{\gamma}) =  \mv{\Sigma}_{X_2 X_1}^{tap}(\mv{\Sigma}_{X_1}^{tap}+\mv{\Sigma}_{\mathcal{E}})^{-1}\mv{Y}\notag
\end{align*}
where the element on row $i$ and column $j$ in $\mv{\Sigma}_{X_2 X_1}^{tap}$
and $\mv{\Sigma}_{X_1}^{tap}$ are given by $r_{tap}(\hat{\mv{s}}_i,\mv{s}_j)$ and $r_{tap}(\mv{s}_i,\mv{s}_j)$ respectively. Since the tapered covariance is zero for lags larger than the taper range, $\theta$, many of the elements in
$\mv{\Sigma}_{X_1}^{tap}$ will be zero. Thus, the three step approach used for the wavelet Markov approximations can be used to solve the system $\mv{u} =
(\mv{\Sigma}_{X_1}^{tap}+\mv{\Sigma}_{\mathcal{E}})^{-1}\mv{Y}$
efficiently. Since the number of non-zero elements for row $i$ in
$\mv{\Sigma}_{X_1}^{tap}$ is determined by the number of measurement
locations at a distance smaller than $\theta$ from location $\mv{s}_i$, the
computational cost is determined both by the taper range and the
spacing of the observations. Thus, if the measurements are irregularly
spaced, it is hard to get a precise estimate of the computational
cost. However, for given measurement locations, the taper range can be
chosen such that the average number of neighbors to the measurement
locations is some fixed number $k_{\theta}$. The cost for the Cholesky
factorization is then similar to the cost for a GMRF with $m$
nodes and a neighborhood size $k_{\theta}$.

\subsection{Covariance approximation}\label{sec:covcomp}
For practical applications of any of the approximation methods discussed here, one is often mostly interested in producing kriging predictions which are close to the optimal predictions. The error one makes in the kriging prediction is closely related to the methods ability to reproduce the true Mat\'{e}rn covariance function. There are many different wavelet bases one could consider using in the Markov approximation method, and before we consider the kriging problem we will in this section compare some of these bases with respect to their ability to reproduce the Mat\'{e}rn covariance function so that we can choose only a few of the best methods to compare in the next section. As a reference, we also include the process convolution approximation in this comparison.

A natural measure of the error in the covariance approximation is a
standardized $L^2$ norm of the difference between the true-, and approximate
covariance functions, 
\begin{equation}\label{paperB:eq:L2norm}
  \ep_r(\mv{s}) = \frac{\int (r(\mv{s},\mv{u})-\hat{r}(\mv{s},\mv{u}))^2 \md \mv{u}}{\int r(\mv{s},\mv{u})^2 \md \mv{u}}.
\end{equation}
Note here that the true covariance function $r(\mv{s},\mv{u})$ is stationary and isotropic, while the approximate covariance function $\hat{r}(\mv{s},\mv{u})$, for the basis expansion \eqref{paperB:lowrank_proc}, generally is not. For the wavelet approximations and the process convolutions, $\ep_r$ is periodic in $\mv{s}$ since the approximation error in general is smaller where the basis functions are centered, and we therefore use the mean value of $\ep(\mv{s})$ over the studied region as a measure of the covariance error. 

We use the different methods to approximate the covariance function for a Mat\'{e}rn field on the square $[0, 10]\times[0, 10]$ in $\R^2$. The computational complexity for the kriging predictions depend on the number of basis functions, $n$, used in the approximations. For the Markov approximated spline bases and the Daubechies 3 basis, this complexity is $O(n^{3/2})$ whereas it is $O(n^3)$ for the spline bases if the Markov approximation is not used and for the process convolution method. We therefore use $100^2$ basis functions for the $O(n^{3/2})$ methods and $100$ basis functions for the other methods to get the covariance error for the methods when the computational cost is approximately equal. 

Figure \ref{paperB:fig:cov_err} shows the covariance error for the different methods as functions of the approximate range, $\kappa^{-1}\sqrt{8\nu}$, of the true covariance function for three different values of $\nu$. There are
several things to note in this figure:
\begin{figure}[t]
\begin{center}
%
%
\begin{psfrags}%
\psfragscanon%
%
\psfrag{s03}[b][b]{\setlength{\tabcolsep}{0pt}\begin{tabular}{c}$\nu = 1$\end{tabular}}%
\psfrag{s04}[t][t]{\setlength{\tabcolsep}{0pt}\begin{tabular}{c}Correlation range\end{tabular}}%
%
\psfrag{x01}[t][t]{0}%
\psfrag{x02}[t][t]{0.2}%
\psfrag{x03}[t][t]{0.4}%
\psfrag{x04}[t][t]{0.6}%
\psfrag{x05}[t][t]{0.8}%
\psfrag{x06}[t][t]{1}%
\psfrag{x07}[t][t]{1.2}%
\psfrag{x08}[t][t]{1.4}%
\psfrag{x09}[t][t]{1.6}%
\psfrag{x10}[t][t]{1.8}%
\psfrag{x11}[t][t]{2}%
%
\psfrag{v01}[r][r]{$10^{-7}$}%
\psfrag{v02}[r][r]{$10^{-6}$}%
\psfrag{v03}[r][r]{$10^{-5}$}%
\psfrag{v04}[r][r]{$10^{-4}$}%
\psfrag{v05}[r][r]{$10^{-3}$}%
\psfrag{v06}[r][r]{$10^{-2}$}%
%
\resizebox{8cm}{!}{\includegraphics{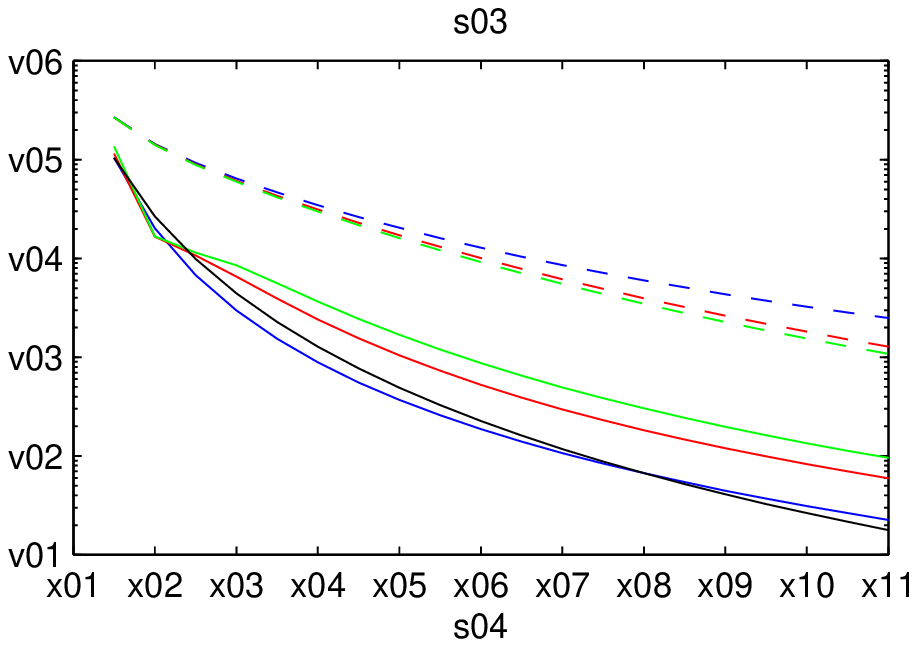}}%
\end{psfrags}%
%
%
%
%
\begin{psfrags}%
\psfragscanon%
%
\psfrag{s03}[b][b]{\setlength{\tabcolsep}{0pt}\begin{tabular}{c}$\nu = 2$\end{tabular}}%
\psfrag{s04}[t][t]{\setlength{\tabcolsep}{0pt}\begin{tabular}{c}Correlation range\end{tabular}}%
%
\psfrag{x01}[t][t]{0}%
\psfrag{x02}[t][t]{0.2}%
\psfrag{x03}[t][t]{0.4}%
\psfrag{x04}[t][t]{0.6}%
\psfrag{x05}[t][t]{0.8}%
\psfrag{x06}[t][t]{1}%
\psfrag{x07}[t][t]{1.2}%
\psfrag{x08}[t][t]{1.4}%
\psfrag{x09}[t][t]{1.6}%
\psfrag{x10}[t][t]{1.8}%
\psfrag{x11}[t][t]{2}%
%
\psfrag{v01}[r][r]{$10^{-8}$}%
\psfrag{v02}[r][r]{$10^{-7}$}%
\psfrag{v03}[r][r]{$10^{-6}$}%
\psfrag{v04}[r][r]{$10^{-5}$}%
\psfrag{v05}[r][r]{$10^{-4}$}%
\psfrag{v06}[r][r]{$10^{-3}$}%
\psfrag{v07}[r][r]{$10^{-2}$}%
%
\resizebox{8cm}{!}{\includegraphics{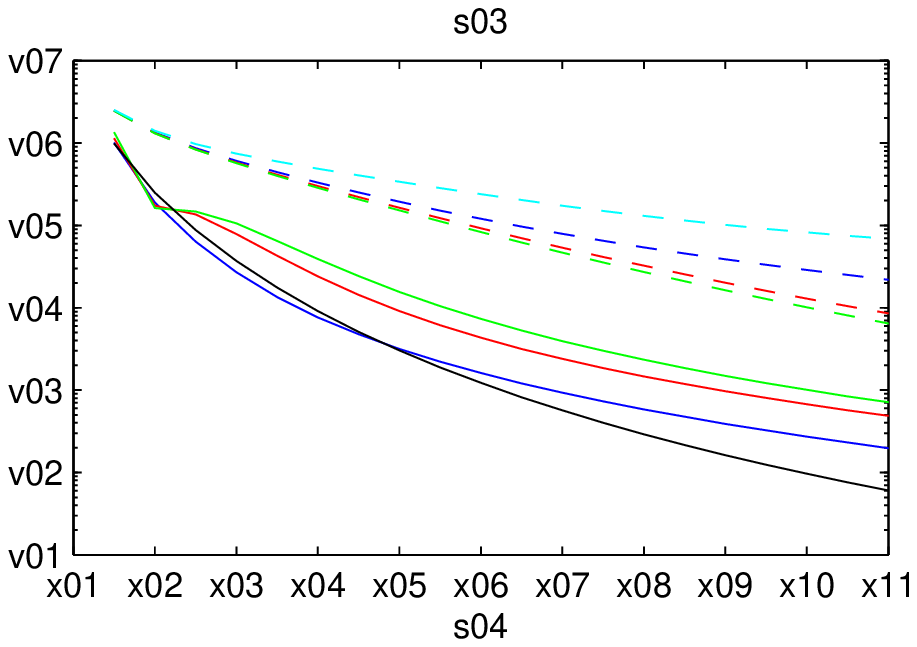}}%
\end{psfrags}%
%
\\%
%
%
\begin{psfrags}%
\psfragscanon%
%
\psfrag{s03}[b][b]{\setlength{\tabcolsep}{0pt}\begin{tabular}{c}$\nu = 3$\end{tabular}}%
\psfrag{s04}[t][t]{\setlength{\tabcolsep}{0pt}\begin{tabular}{c}Correlation range\end{tabular}}%
%
\psfrag{x01}[t][t]{0}%
\psfrag{x02}[t][t]{0.2}%
\psfrag{x03}[t][t]{0.4}%
\psfrag{x04}[t][t]{0.6}%
\psfrag{x05}[t][t]{0.8}%
\psfrag{x06}[t][t]{1}%
\psfrag{x07}[t][t]{1.2}%
\psfrag{x08}[t][t]{1.4}%
\psfrag{x09}[t][t]{1.6}%
\psfrag{x10}[t][t]{1.8}%
\psfrag{x11}[t][t]{2}%
%
\psfrag{v01}[r][r]{$10^{-8}$}%
\psfrag{v02}[r][r]{$10^{-7}$}%
\psfrag{v03}[r][r]{$10^{-6}$}%
\psfrag{v04}[r][r]{$10^{-5}$}%
\psfrag{v05}[r][r]{$10^{-4}$}%
\psfrag{v06}[r][r]{$10^{-3}$}%
\psfrag{v07}[r][r]{$10^{-2}$}%
%
\resizebox{8cm}{!}{\includegraphics{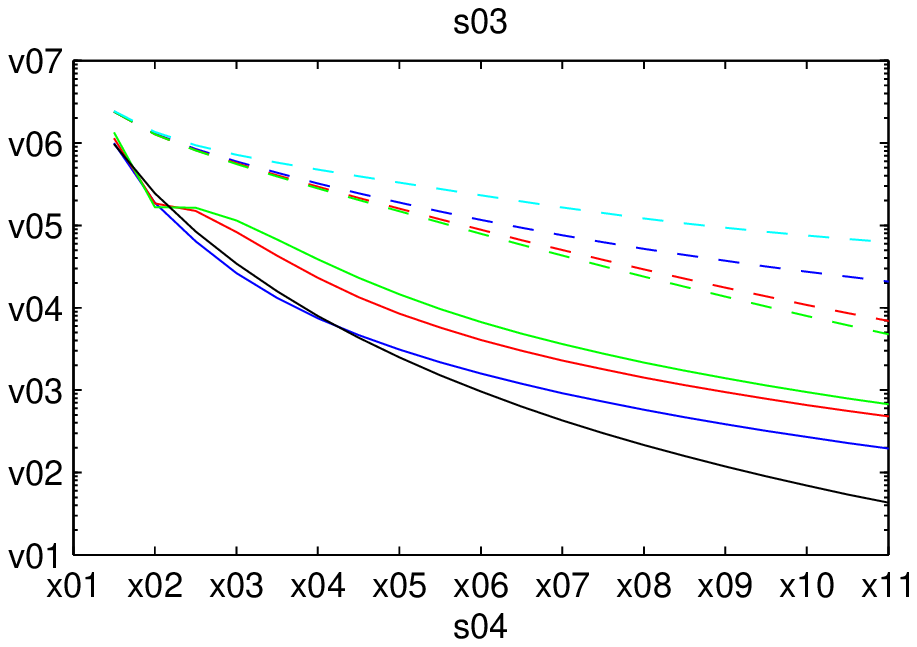}}%
\end{psfrags}%
%
%
%
%
\begin{psfrags}%
\psfragscanon%
%
\psfrag{s10}[][]{\setlength{\tabcolsep}{0pt}\begin{tabular}{c} \end{tabular}}%
\psfrag{s11}[][]{\setlength{\tabcolsep}{0pt}\begin{tabular}{c} \end{tabular}}%
\psfrag{s12}[l][l]{Daubechies 3}%
\psfrag{s13}[l][l]{1st order spline}%
\psfrag{s14}[l][l]{2nd order spline}%
\psfrag{s15}[l][l]{3rd order spline}%
\psfrag{s16}[l][l]{Process convolution}%
\psfrag{s17}[l][l]{Approx 1st order spline}%
\psfrag{s18}[l][l]{Approx 2nd order spline}%
\psfrag{s19}[l][l]{Approx 3rd order spline}%
\psfrag{s20}[l][l]{Daubechies 3}%
%
\psfrag{x01}[t][t]{-1}%
\psfrag{x02}[t][t]{-0.5}%
\psfrag{x03}[t][t]{0}%
\psfrag{x04}[t][t]{0.5}%
\psfrag{x05}[t][t]{1}%
%
\psfrag{v01}[r][r]{-1}%
\psfrag{v02}[r][r]{-0.5}%
\psfrag{v03}[r][r]{0}%
\psfrag{v04}[r][r]{0.5}%
\psfrag{v05}[r][r]{1}%
%
\resizebox{8cm}{!}{\includegraphics{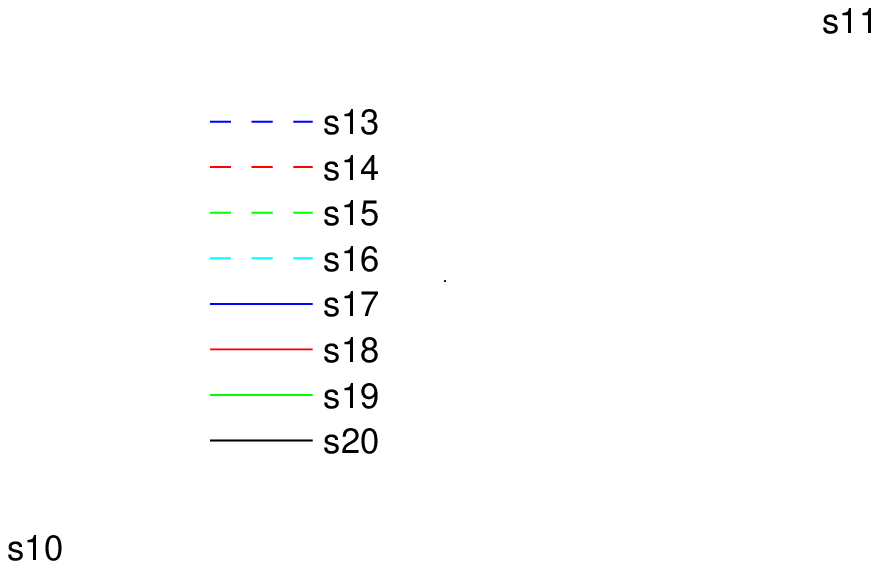}}%
\end{psfrags}%
%
%
\end{center}
\caption{Numeric approximations of the $L^2$-norm \eqref{paperB:eq:L2norm}
  shown as a function of approximate range for different values of
  $\nu$ and different bases in $\R^2$. In this figure, $100^2$ basis
  functions are used for the bases with Markov structure (solid lines), and
  $100$ basis functions are used for the other bases
  (dashed lines). This gives approximately the same computational
  complexity for kriging prediction.}
\label{paperB:fig:cov_err}
\end{figure}

\begin{enumerate}
\item The covariance error decreases for all methods as the range of the true covariance function increases. This is not surprising since the error will be small if the distance between the basis functions (which is kept fixed) is small compared to the true range.  
\item The solid lines correspond to Markov approximations, which have computational complexity $\Ordo(n^{3/2})$ for calculating the kriging
  predictor, and the approximations with computational complexity $\Ordo(n^{3})$ have dashed lines in the figure. 
\item  There is no convolution kernel estimate for $\nu=1$ since the
  convolution kernel has a singularity in zero in this case. For the other cases, the locations $\{u_j\}$ for the kernel basis functions were placed on a regular $10\times 10$ lattice in the region.   
\item  The error one makes by the Markov approximation of the spline bases becomes larger for increasing order of the splines. Note that the third order spline basis is best without the approximation whereas the first order spline basis is best if the Markov approximation is used.
\end{enumerate}

It is clear from the figure that the Markov approximations have a much lower covariance error for the same computational complexity. Among these, the Daubechies 3 basis is best for large ranges whereas the Markov approximated first order spline basis is best for short ranges. The higher order spline bases have larger covariance errors so we from now on focus on the first order spline basis and the Daubechies 3 basis.

\subsection{Spatial prediction}\label{sec:krigingcomp}
In the previous section, several bases were compared with respect to
their ability to approximate the true covariance function when used in an approximation on the form \eqref{paperB:lowrank_proc} of a Gaussian Mat\'{e}rn field. The comparison showed that the Daubechies 3 (DB3) basis and the
Markov approximated linear spline (S1) basis are most accurate for a given
computational complexity. In this section, the spatial prediction
errors for these two wavelet Markov approximations are compared with the process convolution method and the covariance tapering method. In the comparisons, note that the S1 basis is essentially of the same type of piecewise linear basis as used in \cite{lindgren10}, so that the results here also apply to that paper.

\subsubsection*{Simulation setup}
Let $X(\mv{s})$ be a Mat\'{e}rn field with shape parameter $\nu$ (chosen later as $1$, $2$, or $3$) and approximate correlation range $r$ (later varied between $0.1$ and $4$). The range $r$ determines $\kappa$ through the relation $\kappa = \sqrt{8\nu}r^{-1}$ and the variance parameter $\phi=4\pi\Gamma(\nu+1)\kappa^{\nu}\Gamma(\nu)^{-1}$ is chosen such that the variance of $X(\mv{s})$ is $1$. We measure $X$ at $5000$ measurement locations chosen at random from a uniform distribution on the square $[0,5]\times[0,5]$ in $\R^2$ using the measurement equation \eqref{paperB:eq:measurement}, where $\mathcal{E}(\mv{s})$ is Gaussian white noise uncorrelated with $X$ with standard deviation $\sigma=0.01$. 

Given these measurements, spatial prediction of $X$ to all locations on a $70 \times 70$ lattice of equally spaced points in the square is performed using the optimal kriging predictor, the wavelet Markov approximations, the process convolution method, and the covariance tapering method. For each approximate method, the sum of squared differences between the optimal kriging prediction and the approximate method's kriging prediction is used as a measure of
kriging error.

We compare the methods for $\nu=1,2,3$, and for each $\nu$ we test $40$ different ranges varied between $0.1$ and $4$ in steps of $0.1$. For a given $\nu$ and a given range, $20$ data sets are simulated and the average kriging error is calculated for each method based on these data sets. 

\subsubsection*{Choosing the number of basis functions}
To obtain a fair comparison between the different methods, the number of basis functions for each method should be chosen such that the computation time for the kriging prediction is equal for the different methods. The computations needed for calculating the prediction can be divided into three main steps as follows
\begin{description}
\item[Step 1.] Build all matrices except $\mv{M}$ in step 3 necessary to calculate the kriging predictor. 
\item [Step 2.] Solve the matrix inverse problem for the given method:
\begin{align*}
&\mbox{S1, DB3 and Process convolution:} & &\mv{u} = (\mv{\Sigma}_w^{-1} + \Basis_1^{\trsp}\mv{\Sigma}_{\mathcal{E}}^{-1}\Basis_1)^{-1}\Basis_1\mv{\Sigma}_{\mathcal{E}}^{-1}\mv{Y},\\
&\mbox{Tapering:} & &\mv{u} = \left(\mv{\Sigma}_{X_1}^{tap}+\mv{\Sigma}_{\mathcal{E}}\right)^{-1}\mv{Y},\\
&\mbox{Optimal kriging:} & &\mv{u} = \left(\mv{\Sigma}_{X_1}+\mv{\Sigma}_{\mathcal{E}}\right)^{-1}\mv{Y}.
\end{align*}
\item[Step 3.] Depending on which method that is used, build $\mv{M} = \Basis_2$,
  $\mv{M} = \mv{\Sigma}_{X_2 X_1}^{tap}$, or $\mv{M} = \mv{\Sigma}_{X_2 X_1}$ and calculate
  the kriging predictor $\hat{\mv{X}} = \mv{M}\mv{u}$.
\end{description}
For the optimal kriging predictor, and in some cases for the Tapering
method, the matrix $\mv{M}$ cannot be calculated and stored at once
due to memory constraints if the number of measurements is large. 
Each element in $\hat{\mv{X}}$ is then constructed separately as
$\hat{\mv{X}}_i = \mv{M}_i\mv{u}$, where $\mv{M}_i$ is a row in
$\mv{M}$. It is then natural to include the time it takes to build the
rows in $\mv{M}$ in the time it takes to calculate $\hat{\mv{X}}$, which is
the reason for including the time it takes to build $\mv{M}$ in step 3
instead of step 1. 

The computation time for the first step will be very dependent on the actual implementation, and we will therefore focus on the computation time for the last two steps when choosing the number of basis functions. If one only does kriging prediction to a few locations, the second step will dominate the computation time whereas the third step can dominate if kriging is done to many locations. To get results that are easier to interpret, we choose the number of basis functions such that the time for the matrix inverse problem in step 2 is similar for the different methods.    

Now since the computational complexity for step 2 is $O(n^3)$ for the convolution method and $O(n^{3/2})$ for the Markov methods, one would think that if $n$ basis functions are used in the convolution method and $n^2$ basis functions are used for the Markov methods, the computation time would be equal. Unfortunately it is not that simple. If two different methods have computational complexity $O(n^3)$, this means that the computation time scales as $n^3$ when $n$ is increased for both methods; however, the actual computation time for a \emph{fixed} $n$ can be quite different for the two methods. For example, DB3 is approximately 6 times more computationally demanding than S1 for the same number of basis functions. The reason being that the DB3 basis functions have larger support than the S1 basis functions and this cases the matrices $\mv{B}_1$ and $\mv{\Sigma}^{-1}_w$ for DB3 to contain approximately $6$ times as many nonzero elements compared to S1 for the same number of basis functions. However, the relative computation time will scale as $n_1^{3/2}$ if $n_1$ is increased for both methods. 

To get approximately the same computation time for step 2 for the
different approximation methods, the number of basis functions for S1 is fixed to $100^2$. Since DB3 is approximately six times more computationally demanding, the number of basis functions for this method is set to $1600$. As mentioned in \cite{lindgren10}, one should extend the area somewhat to avoid boundary effects from the SPDE formulation used in the Markov methods. We therefore expand the area with two times the range in each direction which results in a slightly higher number of basis functions used in the computations.

The computation time for S1 and DB3 increases if $\nu$ is increased since the precision matrix for the weights contain more nonzero elements for larger values of $\nu$. Therefore we use $625$ basis functions placed on a regular $25 \times 25$ lattice in the kriging area for the convolution method when $\nu=2$ and use $841$ basis functions placed on a regular $29 \times 29$ lattice when $\nu=3$.
For the tapering method we chose the tapering range $\theta$ such that the expected number of measurements within a circle with radius $\theta$ to each kriging location is similar to the number of neighbors to the weights in the S1 method. For $\nu=1$, $\nu=2$, and $\nu=3$ this gives a tapering ranges of $0.4$, $0.55$, and $0.7$ respectively and results in approximately the same number of nonzero elements in the tapered covariance matrix as in the precision matrix $Q$ for the S1 basis. 

\begin{figure}[t]
\begin{center}
%
%
\begin{psfrags}%
\psfragscanon%
%
\psfrag{s03}[b][b]{\setlength{\tabcolsep}{0pt}\begin{tabular}{c}$\nu = 1$\end{tabular}}%
\psfrag{s04}[t][t]{\setlength{\tabcolsep}{0pt}\begin{tabular}{c}Correlation range\end{tabular}}%
%
\psfrag{x01}[t][t]{0}%
\psfrag{x02}[t][t]{0.5}%
\psfrag{x03}[t][t]{1}%
\psfrag{x04}[t][t]{1.5}%
\psfrag{x05}[t][t]{2}%
\psfrag{x06}[t][t]{2.5}%
\psfrag{x07}[t][t]{3}%
\psfrag{x08}[t][t]{3.5}%
\psfrag{x09}[t][t]{4}%
%
\psfrag{v01}[r][r]{$10^{-2}$}%
\psfrag{v02}[r][r]{$10^{-1}$}%
\psfrag{v03}[r][r]{$10^{0}$}%
\psfrag{v04}[r][r]{$10^{1}$}%
\psfrag{v05}[r][r]{$10^{2}$}%
%
\resizebox{8cm}{!}{\includegraphics{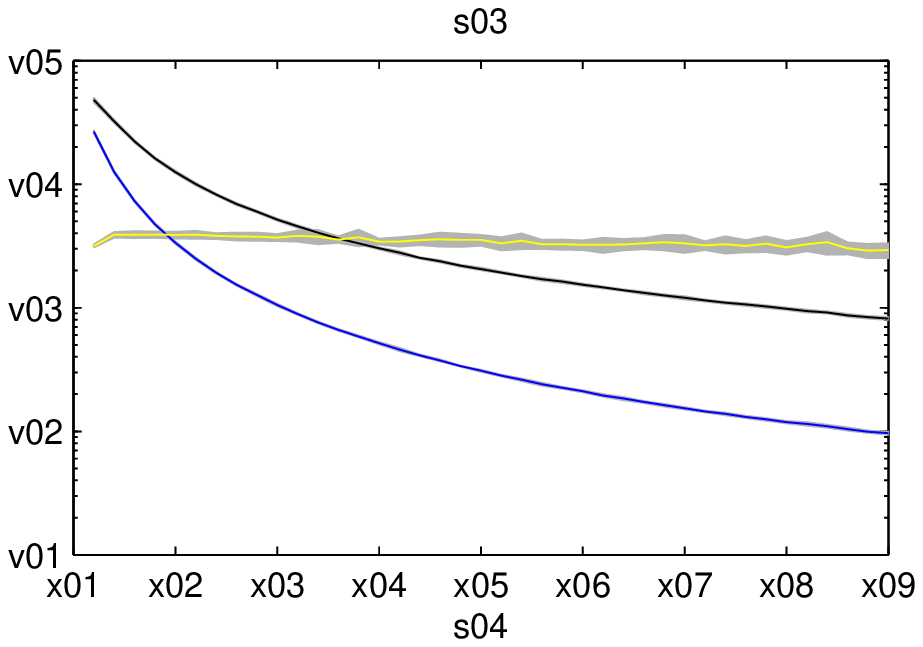}}%
\end{psfrags}%
%
%
%
%
\begin{psfrags}%
\psfragscanon%
%
\psfrag{s03}[b][b]{\setlength{\tabcolsep}{0pt}\begin{tabular}{c}$\nu = 2$\end{tabular}}%
\psfrag{s04}[t][t]{\setlength{\tabcolsep}{0pt}\begin{tabular}{c}Correlation range\end{tabular}}%
%
\psfrag{x01}[t][t]{0}%
\psfrag{x02}[t][t]{0.5}%
\psfrag{x03}[t][t]{1}%
\psfrag{x04}[t][t]{1.5}%
\psfrag{x05}[t][t]{2}%
\psfrag{x06}[t][t]{2.5}%
\psfrag{x07}[t][t]{3}%
\psfrag{x08}[t][t]{3.5}%
\psfrag{x09}[t][t]{4}%
%
\psfrag{v01}[r][r]{$10^{-2}$}%
\psfrag{v02}[r][r]{$10^{-1}$}%
\psfrag{v03}[r][r]{$10^{0}$}%
\psfrag{v04}[r][r]{$10^{1}$}%
\psfrag{v05}[r][r]{$10^{2}$}%
\psfrag{v06}[r][r]{$10^{3}$}%
%
\resizebox{8cm}{!}{\includegraphics{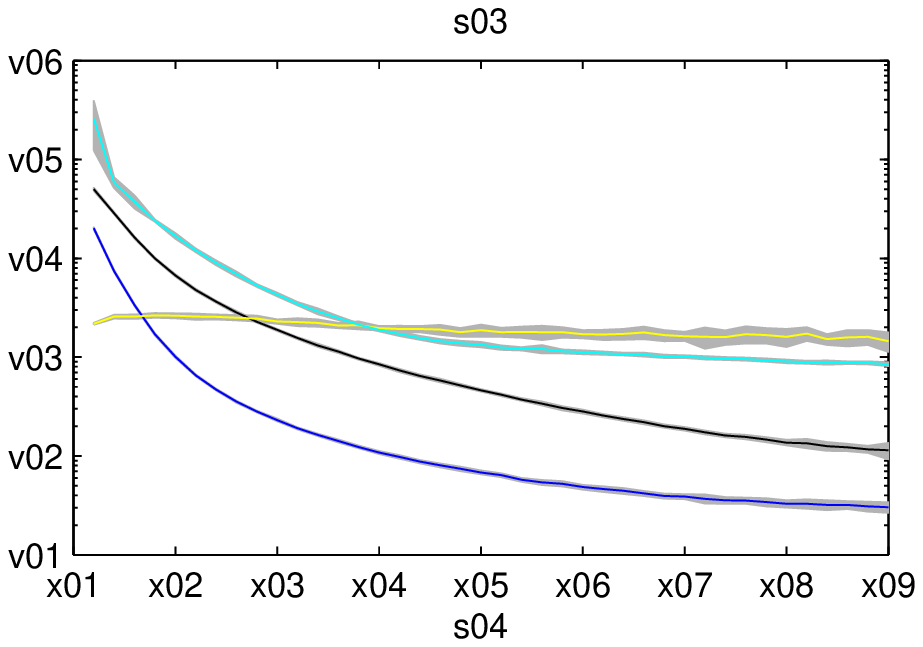}}%
\end{psfrags}%
%
\\%
%
%
\begin{psfrags}%
\psfragscanon%
%
\psfrag{s03}[b][b]{\setlength{\tabcolsep}{0pt}\begin{tabular}{c}$\nu = 3$\end{tabular}}%
\psfrag{s04}[t][t]{\setlength{\tabcolsep}{0pt}\begin{tabular}{c}Correlation range\end{tabular}}%
%
\psfrag{x01}[t][t]{0}%
\psfrag{x02}[t][t]{0.5}%
\psfrag{x03}[t][t]{1}%
\psfrag{x04}[t][t]{1.5}%
\psfrag{x05}[t][t]{2}%
\psfrag{x06}[t][t]{2.5}%
\psfrag{x07}[t][t]{3}%
\psfrag{x08}[t][t]{3.5}%
\psfrag{x09}[t][t]{4}%
%
\psfrag{v01}[r][r]{$10^{-2}$}%
\psfrag{v02}[r][r]{$10^{-1}$}%
\psfrag{v03}[r][r]{$10^{0}$}%
\psfrag{v04}[r][r]{$10^{1}$}%
\psfrag{v05}[r][r]{$10^{2}$}%
\psfrag{v06}[r][r]{$10^{3}$}%
%
\resizebox{8cm}{!}{\includegraphics{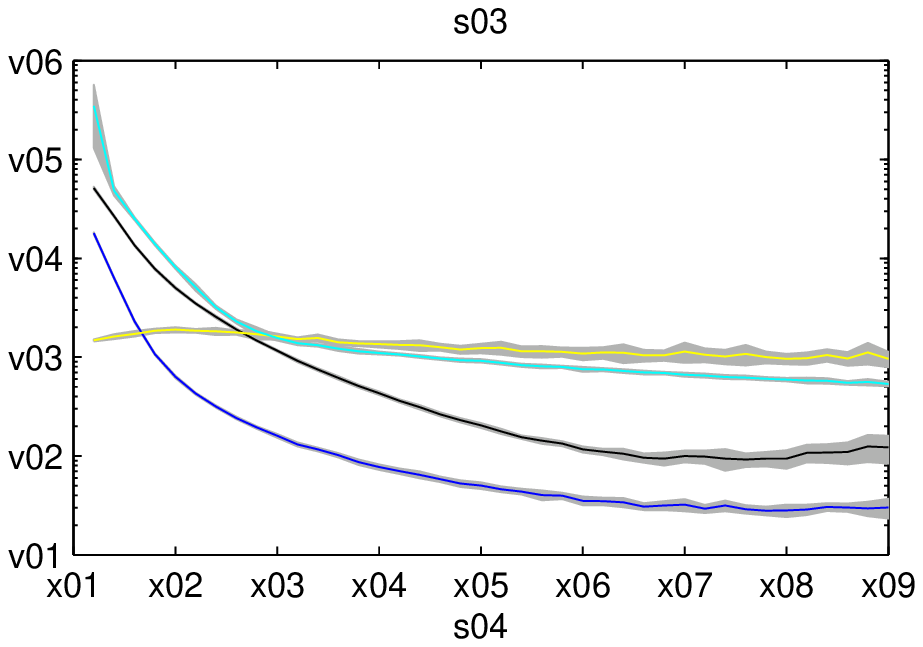}}%
\end{psfrags}%
%

%
%
\begin{psfrags}%
\psfragscanon%
%
\psfrag{s10}[][]{\setlength{\tabcolsep}{0pt}\begin{tabular}{c} \end{tabular}}%
\psfrag{s11}[][]{\setlength{\tabcolsep}{0pt}\begin{tabular}{c} \end{tabular}}%
\psfrag{s12}[l][l]{Tapering}%
\psfrag{s13}[l][l]{S1}%
\psfrag{s14}[l][l]{DB3}%
\psfrag{s15}[l][l]{Process convolution}%
\psfrag{s16}[l][l]{Tapering}%
%
\psfrag{x01}[t][t]{-1}%
\psfrag{x02}[t][t]{-0.5}%
\psfrag{x03}[t][t]{0}%
\psfrag{x04}[t][t]{0.5}%
\psfrag{x05}[t][t]{1}%
%
\psfrag{v01}[r][r]{-1}%
\psfrag{v02}[r][r]{-0.5}%
\psfrag{v03}[r][r]{0}%
\psfrag{v04}[r][r]{0.5}%
\psfrag{v05}[r][r]{1}%
%
\resizebox{8cm}{!}{\includegraphics{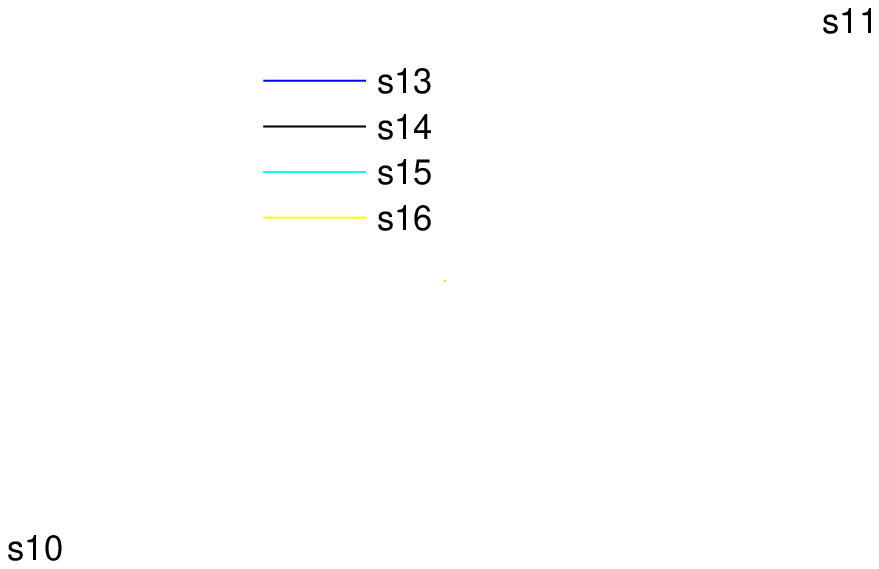}}%
\end{psfrags}%
%

\end{center}
\caption{Kriging errors for the different methods as a functions of the true covariance function's range. For each range, the values are calculated as the mean of $20$ simulations. The lower limit of the bands around the curves are the estimate minus the standard deviation of the samples, and the upper limit is the estimate plus the standard deviation.}
\label{paperB:fig:kriging_err}
\end{figure}

\subsubsection*{Results}
\begin{figure}[t]
\begin{center}
\parbox[b]{4cm}{\centering Optimal prediction \\ 
\includegraphics[width=4cm]{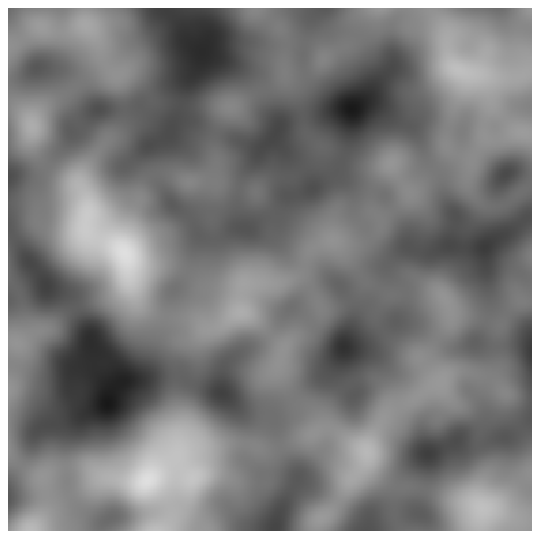}}%
\parbox[b]{4cm}{\centering S1 basis \\
\includegraphics[width=4cm]{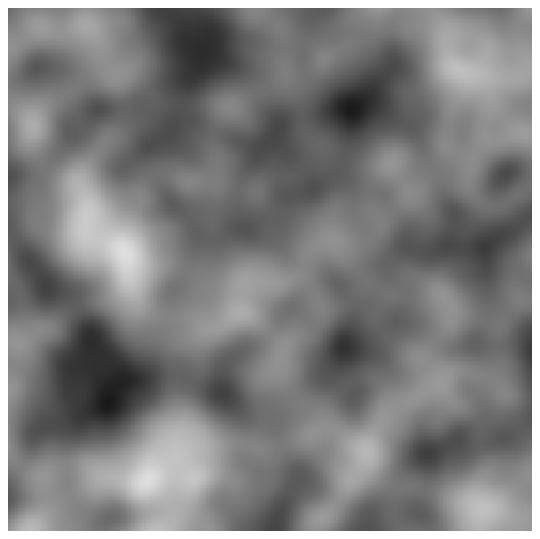}}
\parbox[b]{4cm}{\centering Convolution basis \\
\includegraphics[width=4cm]{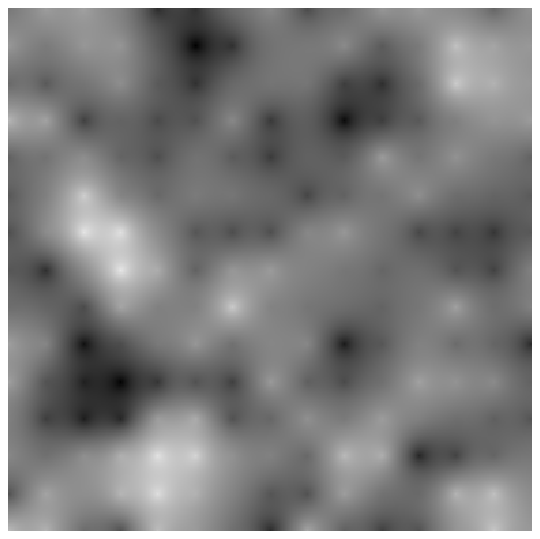}}%
\parbox[b]{4cm}{\centering Tapered covariance \\
\includegraphics[width=4cm]{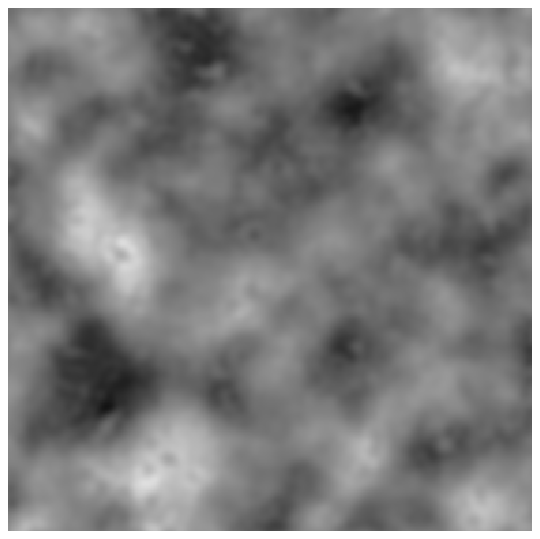}}%
\vspace{-0.5cm}
\end{center}
\caption{An example of an optimal kriging prediction and predictions using the
  S1 basis, the convolution basis, and a tapered
  covariance when $\nu=2$ and the covariance range is $1$. The predictions are based on $5000$ observations and are calculated for a $200\times 200$ grid in the square $[0,5]\times[0,5]$. The number of basis functions and the tapering range are chosen such that the total time for Step 2 and Step 3 is approximately equal for the different methods.}
\label{paperB:fig:kriging_ex1}
\end{figure}

In Figure \ref{paperB:fig:kriging_err} can the average kriging errors for the different methods be seen as functions of the true covariance function's approximate range $r$. The values for a given $\nu$ and $r$ is an average of $20$ simulations.  The convolution kernels are singular if $\nu=1$, so there is no convolution estimate for this case. The tapering estimate is best for short
ranges, which is not surprising since the covariance matrix for the
measurements not is changed much by the tapering if the true range
then is shorter than the tapering range. For larger ranges, however,
the tapering method has a larger error than the other methods.
One reason for this is that the tapered covariance function is very different
from the true covariance function if the true range is much larger than the tapering range. Another reason is that the prediction for all locations that do not have any measurements closer than the tapering range is zero in the tapering method. The convolution method has a similar problem if the effective range of the basis functions is smaller than the distance between the basis functions. In this case, the estimates for all locations that are not close to the center of some basis function have a large bias towards zero. These problems can clearly be seen in Figure \ref{paperB:fig:kriging_ex1}, where the optimal kriging
prediction, and the predictions for S1, the tapering method, and the convolution method, are shown for an example where $\nu=2$ and the range is $1$.

\begin{table}[t]
  \begin{center}
\begin{tabular}{|l| l l | l l | l l | l l| l l|}
\multicolumn{11}{c}{$\nu=1$}\\
\hline
& \multicolumn{2}{c|}{Optimal} & \multicolumn{2}{c|}{DB3} & \multicolumn{2}{c|}{S1} & \multicolumn{2}{c|}{Conv.} & \multicolumn{2}{c|}{Taper}  \\
\hline
Step 1 & $37.68$ & $(6.357)$ & $0.490$ & $(0.049)$ & $0.423$ & $(0.027)$ & $-$ & $-$ & $2.771$ & $(0.191)$\\
Step 2 & $5.074$ & $(0.277)$ & $0.113$ & $(0.014)$ & $0.088$ & $(0.007)$ & $-$ & $-$ & $0.117$ & $(0.010)$\\
Step 3 & $36.48$ & $(6.231)$ & $0.293$ & $(0.026)$ & $0.248$ & $(0.018)$ & $-$ & $-$ & $2.051$ & $(0.127)$\\
\hline
Total & $79.23$ & $(8.906)$ & $0.896$ & $(0.057)$ & $0.759$ & $(0.033)$ & $-$  &  $-$ & $4.939$ & $(0.229)$\\
\hline
\multicolumn{11}{c}{$\nu=2$}\\
\hline
Step 1 & $36.19$ & $(6.965)$ & $0.600$ & $(0.090)$ & $0.489$ & $(0.055)$ & $0.961$ & $(0.027)$ & $4.184$ & $(1.523)$\\
Step 2 & $5.327$ & $(0.529)$ & $0.228$ & $(0.039)$ & $0.203$ & $(0.025)$ & $0.217$ & $(0.019)$ & $0.247$ & $(0.028)$\\
Step 3 & $34.94$ & $(6.695)$ & $0.310$ & $(0.049)$ & $0.260$ & $(0.036)$ & $0.942$ & $(0.027)$ & $3.319$ & $(0.251)$\\
\hline
Total & $76.45$ & $(9.675)$ & $1.138$ & $(0.110)$ & $0.951$ & $(0.070)$ & $2.120$ & $(0.043)$ & $7.750$ & $(1.543)$\\
\hline
\multicolumn{11}{c}{$\nu=3$}\\
\hline
Step 1 & $42.75$ & $(6.572)$ & $0.759$ & $(0.091)$ & $0.569$ & $(0.042)$ & $5.656$ & $(1.094)$ & $6.413$ & $(1.051)$\\
Step 2 & $5.468$ & $(0.380)$ & $0.394$ & $(0.051)$ & $0.377$ & $(0.035)$ & $0.390$ & $(0.024)$ & $0.421$ & $(0.035)$\\
Step 3 & $41.36$ & $(6.440)$ & $0.315$ & $(0.033)$ & $0.266$ & $(0.025)$ & $5.522$ & $(1.078)$ & $5.460$ & $(0.402)$\\
\hline
Total & $89.58$ & $(9.210)$ & $1.468$ & $(0.110)$ & $1.213$ & $(0.060)$ & $11.57$ & $(1.537)$ & $12.30$ & $(1.126)$\\
\hline
\end{tabular}
\end{center}
\caption{Average computation times for the results in Figure \ref{paperB:fig:kriging_err}. The values are based on the $800$ simulations for each value of $\nu$. The standard deviations are shown in the parentheses.}
\label{paperB:tab:kriging}
\end{table}



The computation times for the different methods are shown in Table
\ref{paperB:tab:kriging}. These computation times are obtained using an implementation in Matlab\footnote{implementation available at \url{http://www.maths.lth.se/matstat/staff/bolin/}} on a computer with a 3.33GHz Intel Xeon X5680 processor. As intended, the time for step 2 is similar for the different methods whereas there is a larger difference between the computation time for step 3 because the computation time for the kriging prediction scales differently with the number of kriging locations for the different methods. Note that the wavelet methods are less computationally demanding than the tapering method and the convolution method when doing kriging to many locations. The reason being that the matrix $\mv{M}$ in step 3 can be constructed without having to do costly covariance function evaluations. 

As mentioned previously is the computation time for step 1 very dependent on the actual implementation. However, as for step 3 can the Markov method's matrices be constructed without doing any covariance function evaluations which is the reason for the faster computation time. One thing to note here is that if the parameters are changed (for example when doing parameter estimation), one does not have to construct all matrices again in the Markov methods as one have to do for the other two methods. 

In conclusion we see that S1 is both faster and has a smaller kriging error for all ranges when compared to DB3 and the convolution method and compared to the tapering method it has a smaller kriging error for all but very short ranges. 
Since the tapering methods computational cost varies with the tapering range, we conclude this section with a study of how changing the tapering range changes the results in order to get a better understanding of which method is to prefer when comparing S1 and the tapering method. 

\begin{figure}[t]
\begin{center}
%
%
\begin{psfrags}%
\psfragscanon%
%
\psfrag{s03}[b][b]{\setlength{\tabcolsep}{0pt}\begin{tabular}{c}Kriging error\end{tabular}}%
\psfrag{s04}[t][t]{\setlength{\tabcolsep}{0pt}\begin{tabular}{c}Tapering range\end{tabular}}%
%
\psfrag{x01}[t][t]{0}%
\psfrag{x02}[t][t]{0.2}%
\psfrag{x03}[t][t]{0.4}%
\psfrag{x04}[t][t]{0.6}%
\psfrag{x05}[t][t]{0.8}%
\psfrag{x06}[t][t]{1}%
\psfrag{x07}[t][t]{1.2}%
\psfrag{x08}[t][t]{1.4}%
\psfrag{x09}[t][t]{1.6}%
\psfrag{x10}[t][t]{1.8}%
\psfrag{x11}[t][t]{2}%
%
\psfrag{v01}[r][r]{$10^{-1}$}%
\psfrag{v02}[r][r]{$10^{0}$}%
\psfrag{v03}[r][r]{$10^{1}$}%
\psfrag{v04}[r][r]{$10^{2}$}%
%
\resizebox{8cm}{!}{\includegraphics{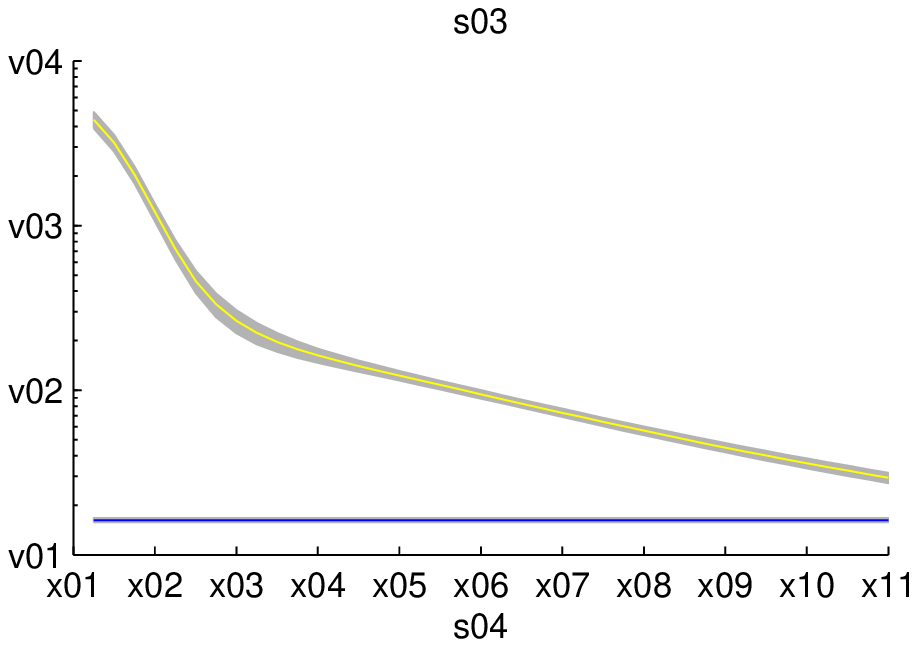}}%
\end{psfrags}%
%
%
%
%
\begin{psfrags}%
\psfragscanon%
%
\psfrag{s02}[b][b]{\setlength{\tabcolsep}{0pt}\begin{tabular}{c}Time\end{tabular}}%
\psfrag{s03}[t][t]{\setlength{\tabcolsep}{0pt}\begin{tabular}{c}Tapering range\end{tabular}}%
\psfrag{s04}[b][b]{\setlength{\tabcolsep}{0pt}\begin{tabular}{c}seconds\end{tabular}}%
%
\psfrag{x01}[t][t]{0}%
\psfrag{x02}[t][t]{0.2}%
\psfrag{x03}[t][t]{0.4}%
\psfrag{x04}[t][t]{0.6}%
\psfrag{x05}[t][t]{0.8}%
\psfrag{x06}[t][t]{1}%
\psfrag{x07}[t][t]{1.2}%
\psfrag{x08}[t][t]{1.4}%
\psfrag{x09}[t][t]{1.6}%
\psfrag{x10}[t][t]{1.8}%
\psfrag{x11}[t][t]{2}%
%
\psfrag{v01}[r][r]{$10^{-3}$}%
\psfrag{v02}[r][r]{$10^{-2}$}%
\psfrag{v03}[r][r]{$10^{-1}$}%
\psfrag{v04}[r][r]{$10^{0}$}%
\psfrag{v05}[r][r]{$10^{1}$}%
\psfrag{v06}[r][r]{$10^{2}$}%
%
\resizebox{8cm}{!}{\includegraphics{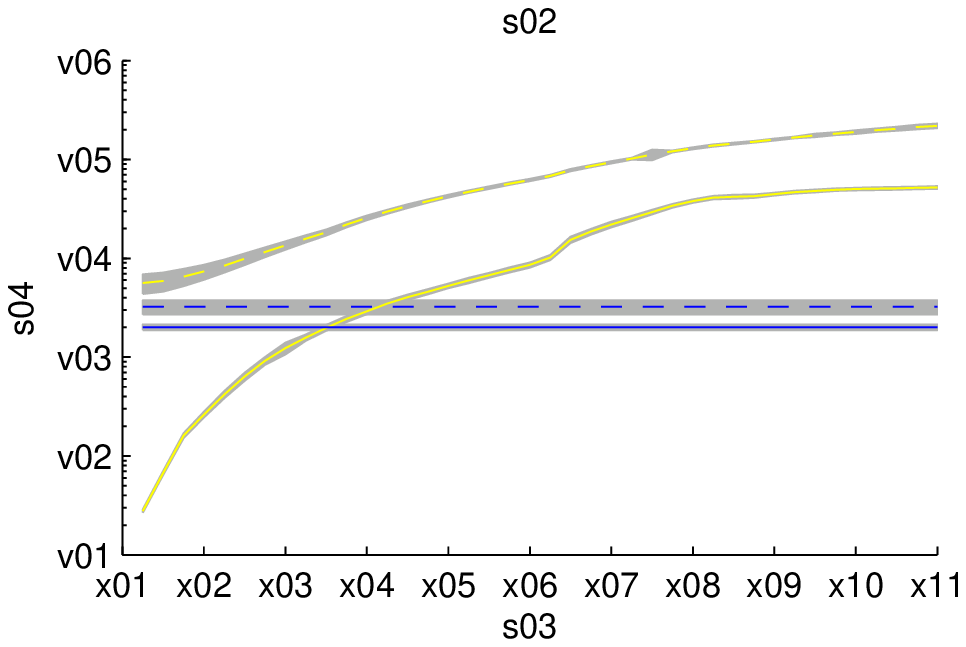}}%
\end{psfrags}%
%
\\%
%
%
\begin{psfrags}%
\psfragscanon%
%
\psfrag{s02}[b][b]{\setlength{\tabcolsep}{0pt}\begin{tabular}{c}Kriging error\end{tabular}}%
\psfrag{s03}[t][t]{\setlength{\tabcolsep}{0pt}\begin{tabular}{c}Tapering range\end{tabular}}%
%
\psfrag{x01}[t][t]{0}%
\psfrag{x02}[t][t]{0.1}%
\psfrag{x03}[t][t]{0.2}%
\psfrag{x04}[t][t]{0.3}%
\psfrag{x05}[t][t]{0.4}%
\psfrag{x06}[t][t]{0.5}%
\psfrag{x07}[t][t]{0.6}%
\psfrag{x08}[t][t]{0.7}%
\psfrag{x09}[t][t]{0.8}%
%
\psfrag{v01}[r][r]{$10^{0}$}%
\psfrag{v02}[r][r]{$10^{1}$}%
\psfrag{v03}[r][r]{$10^{2}$}%
%
\resizebox{8cm}{!}{\includegraphics{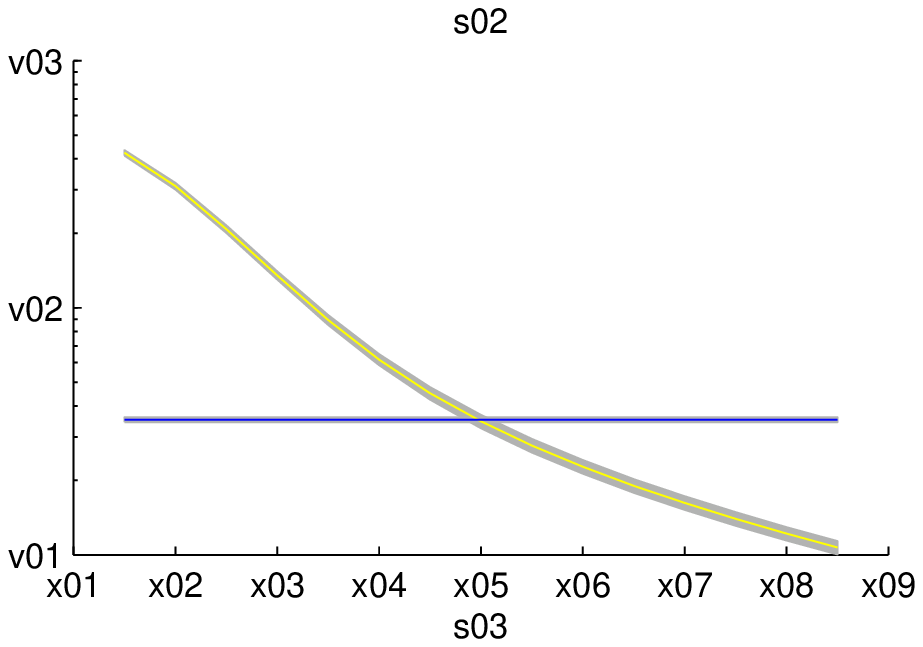}}%
\end{psfrags}%
%
%
%
%
\begin{psfrags}%
\psfragscanon%
%
\psfrag{s02}[b][b]{\setlength{\tabcolsep}{0pt}\begin{tabular}{c}Time\end{tabular}}%
\psfrag{s03}[t][t]{\setlength{\tabcolsep}{0pt}\begin{tabular}{c}Tapering range\end{tabular}}%
\psfrag{s04}[b][b]{\setlength{\tabcolsep}{0pt}\begin{tabular}{c}seconds\end{tabular}}%
%
\psfrag{x01}[t][t]{0}%
\psfrag{x02}[t][t]{0.1}%
\psfrag{x03}[t][t]{0.2}%
\psfrag{x04}[t][t]{0.3}%
\psfrag{x05}[t][t]{0.4}%
\psfrag{x06}[t][t]{0.5}%
\psfrag{x07}[t][t]{0.6}%
\psfrag{x08}[t][t]{0.7}%
\psfrag{x09}[t][t]{0.8}%
%
\psfrag{v01}[r][r]{$10^{-3}$}%
\psfrag{v02}[r][r]{$10^{-2}$}%
\psfrag{v03}[r][r]{$10^{-1}$}%
\psfrag{v04}[r][r]{$10^{0}$}%
\psfrag{v05}[r][r]{$10^{1}$}%
%
\resizebox{8cm}{!}{\includegraphics{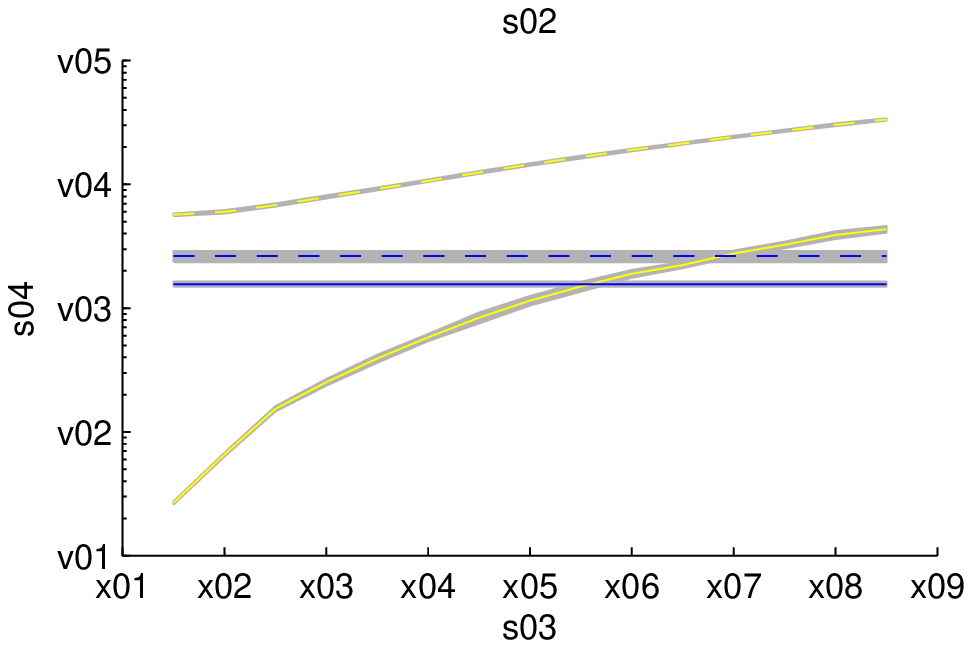}}%
\end{psfrags}%
%

\end{center}
\caption{The computation time for step 2 (right) and the kriging
  errors (left) for the covariance tapering method (yellow lines) as a
  function of taper range.  The values for the S1 basis (blue lines)
  is shown for comparison. In the upper panels, the range of the true covariance function is $1$ and in the lower panels the range is $0.25$. The colored lines are averages of $100$ simulations, and the grey bands indicates the standard deviation of these samples. The solid lines in the right panels show the computation time for step 2 whereas the dashed lines show the total computation time for step 2 and step 3 when calculating the kriging predictions using the two methods. }
\label{paperB:fig:taper_range}
\end{figure}

\subsection{A study of varying the tapering range}\label{sec:tapercomp}
As shown above is the S1 method to prefer over the DB3 method and the convolution method in all our test cases whereas the tapering method had a smaller kriging error for very short ranges. Since this was done using a fixed tapering range chosen such that the computation time for step 2 would be similar to the other methods we now look at what happens if the tapering range is varied when keeping the true range fixed. 

The setup is the same as in the previous comparison, a Mat\'{e}rn field with $\nu=2$, variance $1$ and an approximate range $r$ is measured at $5000$ randomly chosen locations in a square in $\R^2$. The difference is that we now keep these parameters fixed but instead vary the tapering range from $0.05$ to $2$ in steps of $0.05$. We generate $100$ data sets and calculate the kriging predictions for the S1 method and the tapering method for all values of the tapering range. Based on these $100$ estimates, the average kriging error is calculated for S1 and for each tapering estimate. 

The results can be seen in Figure \ref{paperB:fig:taper_range}. The kriging errors are shown in the left panels and the computation times are shown in the right panels. The blue lines represent the S1 method, which obviously does not depend on the tapering range, and the yellow lines represent the tapering method. In the left panels, the solid lines show the time for step 2 in the computations and the dashed lines show the total time for step 2 and step 3. In the upper two panels, the true range $r$ is $1$, and $100^2$ S1 basis functions are used. In this case, S1 is more accurate than the tapering method for all tapering ranges tested, which is not surprising considering the previous results. In the bottom panels of the figure, the true range $r$ is $0.25$ and $100^2$ S1 basis functions are used. This is a case where the tapering method was more accurate than S1 in the previous study and we see here that the tapering method is more accurate for tapering ranges larger than $0.4$ and that the time for step 2 is smaller for all tapering ranges smaller than $0.46$. Thus, by choosing the tapering range between $0.4$ and $0.46$, the tapering method is more accurate and has a smaller computation time for step 2. 

The accuracy of the tapering method increases if the ratio between the tapering range and the true range is increased, and the computation time depends on what the distance between the measurements is compared to the tapering range. If the distance between the measurements is large, the tapering method is fast, whereas it is slower if the distance is small. Thus, the situation where the tapering method performs the best is when the true covariance range is short compared to the distance between the measurements. However, also for the case when the true range is small, the total time it takes to calculate the tapering prediction is larger than the time it takes to calculate the S1 prediction unless the number of kriging locations is small.  

In this work, the taper functions that \cite{furrer06} found
to be best for each value of $\nu$ are used, but the results may be improved
by using other taper functions. Changing the taper function will, however, not change the fact that the prediction for all locations that do not have any measurements closer than the tapering range is zero in the tapering method and that the tapered covariance function is very different from the true covariance function if the tapering range is short compared to the true range. Finally, the results for all methods could be improved by finding optimal parameters for the approximate models instead of using the parameters for the true Mat\'{e}rn covariance. For the tapering method, however, \cite{furrer06} found that this only changed the relative accuracy by one or two percent. 

\section{Conclusions}
Because of the increasing number of large environmental data sets,
there is a need for computationally efficient statistical models.  To
be useful for a broad range of practical applications, the models
should contain a wide family of stationary covariance functions, and
be extendable to nonstationary covariance structures, while still
allowing efficient calculations for large problems.

The SPDE formulation of the Mat\'{e}rn family of covariance functions
has these properties, as it can be extended to more general nonstationary spatial models~\citep[see][for details on how this can be done]{bolin09b,lindgren10}, and allows for efficient and
accurate Markov model representations.  In addition, as shown by the
simulation comparisons, these Markov methods are more efficient and
accurate than both the process convolution approach and the covariance
tapering method for approximating Mat\'ern fields.

Depending on the context in which a model is used, different aspects are important to make it computationally efficient. If, for example, the model is used in MCMC simulations, one should be able to generate samples from the model given the parameters efficiently, or if the parameters are estimated in a numerical maximum likelihood procedure, one must be able to evaluate the likelihood efficiently. To limit the scope of this article, only the computational aspects of kriging was considered. However, for practical applications, parameter estimation is likely the most computationally demanding part of the analysis. If maximum likelihood estimation is performed using numerical optimization of the likelihood, matrix inverses similar to the one in Step 2 in Table \ref{paperB:tab:kriging} have to be performed in each iteration of the optimization, and it is therefore important that these inverses can be calculated efficiently. We have not discussed estimation here, but the Markov methods are likely most efficient in this situation as well because these do not require costly Bessel function evaluations when calculating the likelihood. However, this is left for future research to investigate in more detail. An introduction to maximum likelihood estimation using the SPDE formulation can be found in \cite{bolin09b} and \cite{lindgren10}. 

Finally, some relevant methods, such as \cite{cressie08} and \cite{Banerjee08}, was not included in the comparison in order to keep it relatively short and also because they are difficult to compare with the methods discussed here. It would be interesting to include more methods in the comparison, but we leave this for future work. 

\bibliographystyle{model2-names}
\bibliography{Journals_abrv,waverefs,statistics}  

\end{document}